\documentclass[aps,amsmath,onecolumn,amssymb,nofootinbib,floatfix,showpacs]{revtex4}
\usepackage{graphicx}
\usepackage{subfigure}
\usepackage{amsbsy}
\usepackage{bm}
\newcommand{\be}{\begin{equation}}
\newcommand{\ee}{\end{equation}}
\newcommand{\bea}{\begin{eqnarray}}
\newcommand{\eea}{\end{eqnarray}}
\newcommand{\nn}{\nonumber}

\newcommand{\Tr}{{\mathrm{Tr}}}

\begin{document}

\title{Exploring criticality in the QCD-like two quark flavour models}
\author{Vivek Kumar Tiwari}
\email {vivekkrt@gmail.com}
\affiliation{Department of Physics, University of Allahabad, Allahabad 211002, India.}

\date{\today}
\begin{abstract}
The critical end-point (CEP) and critical behaviour in its vicinity, has been explored 
in the two flavour effective chiral models with and without the presence of effective 
Polyakov loop potential.The tricritical point (TCP) in the massless chiral limit has
been located on the phase diagram in the $\mu$ and T plane for the Polyakov loop
extended Quark Meson Model (PQM) and pure Quark Meson (QM) model which become effective
Quantum-chromodynamics (QCD) like models due to the proper accounting of fermionic vacuum
loop contribution in the effective potential.The proximity of the TCP to the QCD critical
end-point (CEP) has been quantified in the phase diagram. The critical region around CEP has
been obtained in the presence as well as the absence of fermionic vacuum loop contribution
in the effective potentials of PQM and QM models. The contours of appropriately normalized
constant quark number susceptibility and scalar susceptibility have been plotted around CEP
in different model scenarios. These contours determine the shape of critical region and facilitate
comparisons in different models such that the influence of fermionic vacuum term and Polyakov loop
potential on the critical behavior around CEP can be ascertained in qualitative as well as 
quantitative terms. Critical exponents resulting from the divergence of quark number 
susceptibility at the CEP, have been calulated and compared with in different model scenarios.
The possible influence of TCP on the critical behavior around CEP, has also been discussed. The 
temperature variation of $\sigma$ and $\pi$ meson masses at $\mu=0$, $\mu=\mu_{CEP}$ and $\mu>\mu_{CEP}$
has been shown and compared with in different model scenarios and the emerging mass degeneration 
trend in the $\sigma$ and $\pi$ meson mass variations has been inferred as the chiral symmetry
restoration takes place at higher temperatures.
\end{abstract}


\maketitle

\section{Introduction}
\label{sec:intr}
	Under the extreme conditions of high temperature and/or density, normal hadronic 
matter undergoes a phase transition, where the individual hadrons dissolve into 
their quark and gluon constituents and produce a collective form of matter 
known as the Quark Gluon Plasma(QGP)\cite{Rischke:03,Ortmanns:96ea,Muller,Svetitsky}.
Study of the different aspects of this phase transition,is a tough and challenging
task because Quantum Chromodynamics(QCD) which is the theory of strong interaction,
becomes nonperturbative in the low energy limit. However the QCD vacuum reveals itself
through the process of spontaneous chiral symmetry breaking and phenomenon of color confinement.

 The QCD Lagrangian is known to have the global
$SU_{L+R}(N_{f}) \times SU_{L-R}(N_{f})$ symmetry for $ N_{f} $ flavours 
of massless quarks. The formation of a chiral condensate
in the low energy hadronic vacuum of QCD, leads to the spontaneous breaking of
the axial (A=R-L) part of this symmetry known as the chiral symmetry 
and one gets $ (N_{f}^2-1) $ massless
Goldstone bosons according to the Goldstone's theorem. Since quarks are not 
massless in real life, chiral symmetry of the QCD lagrangian gets explicitly broken 
and massless modes become pseudo-Goldstone bosons after acquiring mass. Nevertheless,
the observed lightness of pions in nature suggests that we have an approximate 
chiral symmetry for QCD with two falvours of light u and d quarks.
In the opposite limit of infinitely heavy quarks, QCD becomes a pure $SU(N_{c})$
gauge theory which remains invariant under the global $Z(N_{c})$ center symmetry of
the color gauge group. The Center symmetry which is a symmetry of hadronic vacuum, 
gets spontaneously broken in the high temperature/density regime of QGP. 
The expectation value of the Wilson line (Polyakov loop) is related 
to the free energy of a static color charge. It vanishes in the confining phase as 
the quark has infinite free energy and becomes finite in the deconfined phase.
Hence the Polyakov loop serves as the order parameter of the confinement-deconfinement 
phase transition \cite{Polyakov:78plb}. Even though the center symmetry is always broken 
with the inclusion of dynamical quarks in the system, one can regard the 
Polyakov loop as an approximate order parameter because it is a good 
indicator of a rapid crossover in the confinement-deconfinement transition \cite{Pisarski:00prd,Vkt:06}.

  Lattice QCD simulations (see e.g.~\cite{AliKhan:2001ek,Karsch:02,Forcra,Fodor:03,Allton:05,Aoki:06,Karsch:05,
Karsch:07ax,Cheng:06,Cheng:08,Digal:01}) give us important information and insights 
regarding various aspects of the QGP transition,
like the restoration of chiral symmetry in QCD, order of the 
confinement-deconfinement phase transition, richness of the QCD phase 
structure and phase diagram mapping. Since lattice 
calculations are technically involved and various issues are not conclusively settled 
within the lattice community, one resorts to the calculations within the
ambit of phenomenological models developed in terms of effective degrees 
of freedom. These models serve to complement the lattice simulations and 
give much needed insight about the regions of phase diagram inaccessible 
to lattice simulations.

Construction and mapping of the phase diagram in the quark chemical potential and temperature 
plane is the prime challenge before the experimental as well as theoretical QGP community. 
On the temperature axis, the chiral transition 
at zero quark chemical potential with almost physical quark masses, has been well established to 
be a crossover in recent lattice QCD simulations\cite{Aoki:06,Bazavov}. Effective chiral model studies\cite{Wilczek}
predict first order phase transition at lower tempertaures on the chemical potential axis. Thus the existence 
of a critical end point(CEP) has been suggested in the phase diagram based on model studies\cite{Asaka,Barducci,Berges}
together with the inputs from lattice simulations\cite{Forcra,Fodor:03,Allton:05}. The first order transition
line starting from the lowest 
temperarute on the chemical potential axis, terminates at the CEP which is a genuine singularity of the QCD
free energy. Here the phase transition turns second order and its criticality belongs to the three dimensional 
Ising universality class \cite{Hatta,Fujii,KFuku,Sour}. The precise location of the CEP is highly 
sensitive to the value  of the strange quark mass. Lattice QCD predictions at non zero chemical potential are
much more difficult due to the QCD action becoming complex on account 
of the fermion sign problem \cite{Karsch:02}. There is evidence for a CEP at
finite $\mu$\cite{Forcra,Fodor:03} from a Taylor expansion 
of QCD pressure around $\mu= 0$, however in another lattice study, finite 
chemical potential extrapolations provide some limitations and can rule 
out the existence of a CEP for small $\mu$/T ratios\cite{Philip}. In the
chiral limit of zero up and down quark masses, the chiral phase transition 
is of second order at zero $\mu$ and the static critical behavior is
expected to fall in the universality class of the O(4) spin model in 
three dimensions\cite{Wilczek}. Thus the existance of CEP for real life
two flavor QCD implies that two flavor massless QCD has a tricritical
point(TCP) at which the second order O(4) line of critical points ends.

	Experimental signatures encoding the singular behavior of thermodynamic quantities
in the vicinity of critical point have already been suggested\cite{Stephan}. These are 
related to chemical potential and temperature fluctuations in event-by-event fluctuations
of various particle multiplicities\cite{Jeon}. In the centre of mass energy scans, an 
increase and then  a decrease in the number fluctuations of pions and protons should be 
observed as one crosses the critical point. If the signals are not washed out due to the
expansion of the colliding system, the critical point might be located in the phase diagram
by the observation of  nonmonotonic behavior of number fluctuations in its vicinity\cite{EjiK}.
Recently "beam energy scan" program dedicated to the search of critical point has been started 
at the Relativistic Heavy Ion Collider (RHIC, Brookhaven National Laboratory) experiments\cite{Aggar}.
The Compressed Baryonic Matter (CBM) experiment (GSI-Darmstadt) at the facility for Antiproton and Ion Research
(FAIR) and the Nuclotron-Based Ion Collider facility (NICA) at the Joint Institute for 
Nuclear Research (JINR), will also be looking for the signatures of critical end point. 
Characteristic signatures of the conjectured CEP  for experiments have been discussed in 
refs\cite{Beda,Vkok,Luo}.

	Recently, effective chiral models like the linear sigma models(LSM) \cite{Rischke:00,Hatsuda,Chiku,Herpay:05,
Herpay:06,Herpay:07,Fejos},the quark-meson (QM) models(see e.g.\cite{Mocsy:01prc, Schaefer:09,
Andersen,jakobi,mocsy,bj,Schaefer:2006ds,Kovacs:2006ym,Bowman:2008kc,Jakovac:2010uy,koch}), 
Nambu-Jona-Lasinio (NJL) type models \cite{Costa:08, Mocsy:01prc, Kneur,kahara,nickel}, 
were extended to combine the features of confinement-deconfinement transition together with that of
chiral symmetry breaking-restoring phase transition. Chiral order parameter and Polyakov 
loop order parameter got simultaneously coupled to the quark degrees of freedom in these models. 
Thus Polyakov loop augmented PNJL models \cite{Fukushima:04plb,Ratti:06, Ratti:07,Ratti:07npa,Tamal:06,Sasaki:07,Hell:08,
Abuki:08,Ciminale:07,Fu:07,Fukushima:08d77,Fukushima:08d78,Fukushima:09,Ratti:07prd,Costa:09CEP,mats,nonlocal}
,PLSM models and PQM models\cite{Schaefer:07,Schaefer:08ax,Schaefer:09wspax,
Schaefer:09ax,H.mao09,gupta,Marko:2010cd,Skokov:2010sf,Herbst:2010rf}
have facilitated the investigation of the full QCD thermodynamics and phase structure at 
zero and finite quark chemical potential and it has been shown that bulk thermodynamics of the effective 
models agrees well with the lattice QCD data. The issue of location of CEP in phase diagramn
together with the extent of criticality around it, is also being actively pursued in
a variety of  effective model studies\cite{Hatta,Fujii}\cite{Pettini,Halasz,Harada,Brouz,Nonaka,Ohtani,
Sousa}. The critical region around CEP is not point-like but has a much richer
structure. The estimation of the size of critical region is especially important for
future experimental seraches of CEP in heavy-ion collision experiments.
 
		In the no-sea mean-field approximations, an ultraviolet divergent part of the 
fermionic vacuum loop contribution to the grand potential got frequently neglected till recently in the 
QM/PQM model calculations\cite{Mocsy:01prc,Schaefer:2006ds,Schaefer:09,kahara,Bowman:2008kc}. 
Due to this, the phase transition on the temperature axis at $\mu=0$
for two flavour QM model becomes first order in the chiral limit of massless quarks and one does not find TCP 
on the phase diagram. Recently, Skokov et al. in Ref. \cite{Skokov:2010sf} addressed this issue by 
incorporating appropriately renormalized fermionic vacuum fluctuations in the thermodynamic potential of the
QM model at zero chemical potential which becomes an effective QCD-like model because now it can reproduce 
the second order chiral phase transition at $\mu=0$ as expected from the universality arguments\cite{Wilczek} 
for the two massless flavours of QCD. The fermionic vacuum correction and its influence has also been 
investigated in earlier works\cite{Mizher:2010zb,Palhares:2008yq,Fraga:2009pi,Palhares:2010be}.
In a recent work\cite{Vivek:12}, we generalized the proper accounting of renormalized fermionic vacuum 
fluctuation in the two flavour PQM model to the non-zero chemical potentials and found that the position
of CEP shifts to a significantly higher chemical potential in the $\mu$ and T plane of the phase diagram,
due to the influence of fermionic vacuum term in our PQMVT (PQM model with vacuum term) model calculations. 
Very recently, Schaefer et. al.\cite{Schaef:12} worked out the size of critical region around CEP in 
a three flavour (2+1) PQM model where cut off independent renormalization of fermionic vacuum fluctuation
has been considered. They calculated critical exponents and higher order non-gaussian moments to identify 
the fluctuations in particle multiplicities. Since the criticality around CEP is influenced by the presence 
of strange quark, it is important to have a two flavor calculation in the same model in order to faclitate 
the comparion with the corresponding size of critical region and nature of criticality obtained in 2+1 
flavour QM/PQM model studies.

In this paper, we will calculate the phase diagram in the massless chiral limit
and locate the tricritical point (TCP) in the $\mu$ and T plane for the PQMVT and QMVT
(QM model with vacuum term) models which have become QCD-like in the presence of
fermionic vacuum term and yield the second order transition at $\mu=0$ on the temperature axis. Further,
we will be investigating the size and extent of critical region around the CEP in phase diagram calculated in the
two flavour QM /PQM models with and without the effect of fermionic vacuum fluctuations in the grand 
potential. We will be plotting the contours of appropriately normalized constant quark number susceptibility 
and scalar susceptibility around CEP in different model scenarios. In order to investigate the qualitative as 
well as quantitative effect of fermionic vacuum term and Polyakov loop potential, on the critical behavior 
around CEP, we will compare the shape of these contours as obtained in different model calculations.
Further, we compute and compare the critical exponents resulting from the divergence of quark number 
susceptibility at the CEP in different model scenarios. The possible influence of TCP on the critical behavior around CEP,
will also be discussed. Finally, we  plot the temperature variation of $\sigma$ and $\pi$ meson masses at
$\mu=0$, $\mu=\mu_{CEP}$ and $\mu>\mu_{CEP}$ in different model scenarios and compare the emerging 
mass degeneration trend in the $\sigma$ and $\pi$ meson mass variations  as the chiral symmetry
gets restored at higher temperatures.


In the presentation of this paper, we recapitulate the formulation of the two quark flavour PQM
model in Sec.\ref{sec:model}. The thermodynamic grand potential and the choice of the Polyakov loop potential  
has been discussed in subsection \ref{subsec:Plgtp}. In the subsection \ref{subsec:Vterm},we give a brief description 
of the appropriate renormalization of fermionic vacuum loop contribution and explain how the new  model parameters
are obtained in vacuum when renormalized vacuum term is added to the effective potential.
The section \ref{sec:TCPCEP} explores the proximity of QCD tricritical point to the critical end-point
and the detail structure of the phase diagram for the QMVT and PQMVT models where the effect of fermionic 
vacuum term has been taken care of in the QM and PQM models. The structure of the phase diagram for QM and PQM
model and the location of critical end point has also been presented to facilitate the comparison.
The subsection \ref{sec:CTRCRIT} investigates the extent of criticality around CEP where contours of constant 
baryon number susceptibility ratios and  constant scalar susceptibility ratios, have been presented in the 
$\mu$ and T plane and comparison in all the four models QM,PQM, QMVT and PQMVT,have been made. The critical exponents for
the criticality around CEP  in all the four models QM,PQM, QMVT and PQMVT, have been discussed in
the subsection \ref{subsec:Expncrit}. Subsection \ref{subsec:Masscrit}, presents the temperature variation 
of $\sigma$ and $\pi$ meson masses at $\mu=0$, $\mu=\mu_{CEP}$ and $\mu>\mu_{CEP}$. Here we also present a detail
comparison of the emerging mass degeneration trends in the $\sigma$ and $\pi$ meson mass variations in different
model scenarios as the chiral symmetry restoration takes place at higher temperatures. In the end Sec. \ref{sec:smry}
presents summary together with the conclusion. The first and second partial derivatives of $\cal U_{\text{log}}$ and 
$\Omega_{\mathrm{q\bar{q}}}^{\rm T}$ with respect to temperature and chemical potential has been evaluated in
appendix A of Ref. \cite{Vivek:12}.

\section{Model Formulation}
\label{sec:model}

We will be working in the two flavor quark meson linear 
sigma model which has been combined with the Polyakov loop potential
\cite{Schaefer:07}
In this model, quarks coming in two flavor are coupled to the 
$SU_L(2)\times SU_R(2)$ symmetric four mesonic fields $\sigma$ and $\vec\pi$ 
together with spatially constant temporal gauge field represented by Polyakov loop 
potential. Polyakov loop field $\Phi(\vec{x})$ is defined as the 
thermal expectation value of color trace of Wilson loop in temporal 
direction 
\be
\Phi = \frac{1}{N_c}\Tr_c L, \qquad \qquad  \Phi^* = \frac{1}{N_c}\Tr_c L^{\dagger}
\ee

where L(x) is a matrix in the fundamental representation of the 
$SU_c(3)$ color gauge group.
\be
\label{eq:Ploop}
L(\vec{x})=\mathcal{P}\mathrm{exp}\left[i\int_0^{\beta}d \tau
A_0(\vec{x},\tau)\right]
\ee
Here $\mathcal{P}$ is path ordering,  $A_0$ is the temporal component
of Euclidean vector field and $\beta = T^{-1}$ \cite{Polyakov:78plb}.
 
The model Lagrangian is written in terms of quarks, mesons, couplings 
and Polyakov loop potential ${\cal U} \left( \Phi, \Phi^*, T \right)$.

\be
\label{eq:Lag}
{\cal L}_{PQM} = {\cal L}_{QM} - {\cal U} \big( \Phi , \Phi^* , T \big) 
\ee
where the Lagrangian in quark meson linear sigma model
\bea
\label{eq:Lqm}
{\cal L}_{QM} = \bar{q_f} \, \left[i \gamma^\mu D_\mu - g (\sigma + i \gamma_5
  \vec \tau \cdot \vec \pi )\right]\,q_f + {\cal L}_{m}
\eea
The coupling of quarks with the uniform temporal background gauge 
field is effected by the following replacement 
$D_{\mu} = \partial_{\mu} -i A_{\mu}$ 
and  $A_{\mu} = \delta_{\mu 0} A_0$ (Polyakov gauge), where 
$A_{\mu} = g_s A^{a}_{\mu} \lambda^{a}/2$. $g_s$ is the $SU_c(3)$ 
gauge coupling. $\lambda_a$ are Gell-Mann matrices in the color 
space, a runs from $1 \cdots 8$. $q_f=(u,d)^T$ denotes the quarks 
coming in two flavors and three colors. g is the flavor blind 
Yukawa coupling that couples the two flavor of quarks with four
mesons; one scalar ($\sigma, J^{P}=0^{+}$) and three pseudo scalars 
($\vec\pi, J^{P}=0^{-}$).

The quarks have no intrinsic mass but become massive after 
spontaneous chiral symmetry breaking because of non vanishing 
vacuum expectation value of the chiral condensate. The mesonic 
part of the Lagrangian has the following form
\bea  
\label{eq:Lagmes}
{\cal L}_{m} & = &\frac 1 2 (\partial_\mu \sigma)^2+ \frac{ 1}{2}
  (\partial_\mu \vec \pi)^2 - U(\sigma, \vec \pi ) 
\eea

The pure mesonic potential is given by the expression
\be
U(\sigma,\vec{\pi})=\frac{\lambda}{4}\left(\sigma^2+\vec{\pi}^2-v^2\right)^2-h\sigma,
\ee
Here $\lambda$ is quartic coupling of the mesonic fields,
v is the vacuum expectation value of 
scalar field when chiral symmetry is explicitly broken 
and $h$ =$f_{\pi} m_{\pi}^2$ .

\subsection{Polyakov loop potential and thermodynamic grand potential}
\label{subsec:Plgtp}

The effective potential ${\cal U} \left( \Phi, \Phi^*, T \right)$ 
is constructed such that it reproduces thermodynamics of pure glue 
theory on the lattice for temperatures upto about twice the 
deconfinement phase transition temperature. In this work, we are 
using logarithmic form of Polyakov loop effective potential \cite{Ratti:07}. 
The results produced by this potential is known to be fitted well to the
lattice results. This potential is given by the following expression

\bea
\label{eq:logpot}
\frac{{\cal U_{\text{log}}}\left(\Phi,\Phi^*, T \right)}{T^4} &=& -\frac{a\left(T\right)}{2}\Phi^* \Phi +
b(T) \, \mbox{ln}[1-6\Phi^* \Phi \nn \\&&+4(\Phi^{*3}+ \Phi^3)-3(\Phi^* \Phi)^2]
\eea

where the temperature dependent coefficients are as follow

\begin{equation*} 
  a(T) =  a_0 + a_1 \left(\frac{T_0}{T}\right) + a_2 \left(\frac{T_0}{T}\right)^2 \; \; \; 
  b(T) = b_3 \left(\frac{T_0}{T}\right)^3\ .
\end{equation*}

The parameters 
of Eq.(\ref{eq:logpot}) are 
\begin{eqnarray*}
&& a_0 = 3.51\ , \qquad a_1= -2.47\ , \nn \\ 
&& a_2 = 15.2\ ,  \qquad  b_3=-1.75\ 
\end{eqnarray*}

\begin{widetext}

The critical temperature for deconfinement phase transition 
$T_0=270$ MeV is fixed for pure gauge Yang Mills theory.
In the presence of dynamical quarks $T_0$ is directly linked to the
mass-scale $\Lambda_{\rm QCD}$, the parameter which
has a flavor and chemical potential dependence in full dynamical QCD
and $T_0\to T_0(N_f,\mu)$ \cite{Schaefer:07,Herbst:2010rf}. For our 
numerical calculations in this paper, we have taken a fixed  $T_0=208$ for 
two flavours of quarks.

In the mean-field approximation, the thermodynamic grand
potential for the PQM model is given as~\cite{Schaefer:07}

\begin{equation}
  \Omega_{\rm MF}(T,\mu;\sigma,\Phi,\Phi^*)  = {\cal
    U}(T;\Phi,\Phi^*) + U(\sigma ) +
  \Omega_{q\bar{q}} (T,\mu;\sigma,\Phi,\Phi^*). 
\label{Omega_MF}
\end{equation}

Here, we have written the vacuum expectation values $\langle \sigma \rangle = \sigma $ 
and $\langle \vec\pi \rangle =0$

The quark/antiquark  contribution in the presence of Polyakov loop reads

\begin{equation}
\Omega_{q\bar{q}} (T,\mu;\sigma, \Phi,\Phi^*) = \Omega_{q\bar{q}}^{\rm vac}+\Omega_{q\bar{q}}^{\rm T}
=- 2 N_f  \int
\frac{d^3 p}{(2\pi)^3} \left\{
 {N_c E_q} \theta( \Lambda^2 - \vec{p}^{\,2})  + T \Bigl[ \ln g_{q}^{+} + \ln g_{q}^{-} \Bigr]
\right\}
\label{Omega_MF_q}
\end{equation}

The first term of the Eq.~(\ref{Omega_MF_q}) denotes the fermion vacuum
contribution, regularized by the ultraviolet cutoff $\Lambda$.
In the second term $g_{q}^{+}$ and $g_{q}^{-}$ have been defined 
after taking trace over color space.

\bea
\label{eq:gpls} 
  g_{q}^{+} =  \Big[ 1 + 3\Phi e^{ -E_{q}^{+} /T} +3 \Phi^*e^{-2 E_{q}^{+}/T} +e^{-3 E_{q}^{+} /T}\Big] \,
\eea
\bea
\label{eq:gmns} 
  g_{q}^{-} =  \Big[ 1 + 3\Phi^* e^{ -E_{q}^{-} /T} +3 \Phi e^{-2 E_{q}^{-}/T} +e^{-3 E_{q}^{-} /T}\Big] \,
\eea

Here we use the notation E$_{q}^{\pm} =E_q \mp \mu $ and $E_q$ is the
single particle energy of quark/antiquark.
\be 
E_q = \sqrt{p^2 + m{_q}{^2}}
\ee 
where the constituent quark mass $m_q=g\sigma$
is a function of chiral condensate. In vacuum $\sigma(0,0) = \sigma_0 = f_\pi=  93.0 MeV$ 
\end{widetext}

\subsection{The renormalized vacuum term and model parameters}
\label{subsec:Vterm}
The fermion vacuum loop contribution can be obtained by appropriately renormalizing 
the first term of Eq.~(\ref{Omega_MF_q}) using the dimensional regularization scheme,
as done in Ref.\cite{Skokov:2010sf}. A brief description of essential steps is given below.

Fermion vacuum term is just the one-loop zero 
temperature effective potential at lowest order ~\cite{Quiros:1999jp}
\begin{eqnarray}
\label{vt_one_loop}
\Omega_{q\bar{q}}^{\rm vac} &=&   - 2 N_f N_c  \int \frac{d^3
  p}{(2\pi)^3} E_q  \nonumber\\ 
&=&    - 2 N_f N_c  \int \frac{d^4 p}{(2\pi)^4}  \ln(p_0^2+E_{q}^{2})
+ {\rm K}, 
\end{eqnarray}
the infinite constant $K$ is independent of
the fermion mass, hence it is dropped.

The dimensional regularization of Eq.~(\ref{vt_one_loop}) near three 
dimensions, $d=3-2\epsilon$ leads to the potential up to zeroth order in $\epsilon$ 
as given by
\begin{equation}
\Omega_{q\bar{q}}^{\rm vac} =  \frac{N_c N_f}{16 \pi^2} m_q^4 \left\{
  \frac{1}{\epsilon} - \frac{1}{2} 
\left[  -3 + 2 \gamma_E + 4 \ln\left(\frac{m_q}{2\sqrt{\pi} M}\right)
\right] 
\right\},
\label{Omega_DR}
\end{equation}
here $M$ denotes the arbitrary renormalization scale.

The addition of a counter term  $\delta \mathcal{L}$ in the Lagrangian of the QM or PQM model 

\begin{equation}
\delta \mathcal{L} = \frac{N_c N_f}{16 \pi^2} g^4\sigma^{4} \left\{ \frac{1}{\epsilon} - \frac{1}{2}
\left[  -3 + 2 \gamma_E - 4 \ln\left(2\sqrt{\pi}\right)  \right] \right\},
\label{counter}
\end{equation}

gives the renormalized fermion vacuum loop contribution as 
\begin{equation}
\Omega_{q\bar{q}}^{\rm reg} =  -\frac{N_c N_f}{8 \pi^2} m_q^4  \ln\left(\frac{m_q}{ M}\right).
\label{Omega_reg}
\end{equation}

Now the first term of Eq.~(\ref{Omega_MF_q}) which is vacuum contribution 
will be  replaced by the appropriately renormalized fermion vacuum loop contribution
as given in Eq.~(\ref{Omega_reg}).

The relevant part of the effective potential in Eq.~(\ref{Omega_MF}) which will fix 
the value of the parameters $\lambda$ and $v$ in the vacuum at $T=0$ and $\mu=0$ is the
purely $\sigma$ dependent mesonic potential $U(\sigma)$ plus the 
renormalized vacuum term given by Eq.~(\ref{Omega_reg}).
\begin{equation}
\Omega(\sigma) = \Omega_{q\bar{q}}^{\rm reg}+U(\sigma)
 = -\frac{N_c N_f}{8 \pi^2} g^4\sigma^4 \ln\left(\frac{g\sigma}{ M}\right)
- \frac{\lambda v^2}{2}\sigma^2+\frac{\lambda}{4}\sigma^4-h\sigma, 
\label{OmegaZEROT}
\end{equation}

The first derivative of $\Omega(\sigma)$ with respect to $\sigma$ at 
$\sigma=f_\pi$ in the vacuum is put to zero

\begin{equation}
\frac{\partial \Omega_{\rm MF}(0,0;\sigma,\Phi,\Phi^*)}{\partial\sigma}
= \frac{\partial \Omega(\sigma)}{\partial \sigma} =0
\label{DERIZERO}
\end{equation}
The second derivative of $\Omega(\sigma)$ with respect to $\sigma$ at 
$\sigma=f_\pi$ in the vacuum gives the mass of $\sigma$

\begin{equation}
m_\sigma ^2 = \frac{\partial^2 \Omega_{\rm MF}(0,0; f_\pi,\Phi,\Phi^*)}{\partial \sigma^2}
	    = \frac{\partial^2 \Omega(\sigma)}{\partial \sigma^2}
\label{msigma2}
\end{equation}

Solving the equations (\ref{DERIZERO}) and (\ref{msigma2}), we obtain 

\begin{equation}
\lambda = \lambda_s+\frac{N_c N_f}{8 \pi^2} g^4\left[3+4\ln\left(\frac{g f_\pi}{ M}\right)\right]
\label{lambd}
\end{equation}

and 

\begin{equation}
\lambda v^2 = (\lambda v^2)_s+\frac{N_c N_f}{4 \pi^2} g^4\ f_\pi^2 
\label{lambdv2}
\end{equation}

where $\lambda_s$  and  $(\lambda v^2)_s$ are the values of the parameters 
in the pure sigma model 

\begin{equation}
\lambda_s =\frac{m_\sigma^2-m_\pi^2}{2 f_\pi^2}
\label{lambs}
\end{equation}

\begin{equation}
(\lambda v^2)_s  =\frac{m_\sigma^2-3m_\pi^2}{2}
\label{lambv*2s}
\end{equation}

It is evident from the equations (\ref{lambd}) and (\ref{lambdv2}) that the value of
the parameters $\lambda$ and $v^2$ have a logarithmic dependence on the arbitrary 
renormalization scale M. However, when we put the value of $\lambda$ and $\lambda v^2$
in Eq.(\ref{OmegaZEROT}), the M dependence cancels out neatly after the rearrangement 
of terms. Finally we obtain 

\begin{equation}
\Omega(\sigma) =  -\frac{N_c N_f}{8 \pi^2} g^4\sigma^4 \ln\left(\frac{\sigma}{f_\pi}\right)
- \frac{\lambda_r v_r^2}{2}\sigma^2+\frac{\lambda_r}{4}\sigma^4-h\sigma, 
\label{OmegasigF}
\end{equation}

Here, we define $\lambda_r$  and  $\lambda_r v_r^2$ as the values of the parameters 
after proper accounting of the renormalized fermion vacuum contribution.

\begin{equation}
\lambda_r = \lambda_s+\frac{3 N_c N_f}{8 \pi^2} g^4
\label{lambdR}
\end{equation}
and 
\begin{equation}
\lambda_r v_r^2 = (\lambda v^2)_s+\frac{N_c N_f}{4 \pi^2} g^4\ f_\pi^2 
\label{Rlambdv2}
\end{equation}

Now the thermodynamic grand potential for the PQM model in the presence of appropriately 
renormalized fermionic vacuum contribution (PQMVT model) will be written as 

\begin{equation}
\Omega_{\rm MF}(T,\mu;\sigma,\Phi,\Phi^*)  = {\cal
U}(T;\Phi,\Phi^*) + \Omega(\sigma ) + 
\Omega_{q\bar{q}}^{\rm T}(T,\mu;\sigma,\Phi,\Phi^*).
\label{OmegaMFPQMVT}
\end{equation}

Thus in the PQMVT model, One can get the chiral condensate $\sigma$, and the 
Polyakov loop expectation values $\Phi$, $\Phi^*$ by searching the global 
minima of the grand potential in Eq.(\ref{OmegaMFPQMVT}) for a given value of 
temperature T and chemical potential $\mu$

\begin{equation}
  \frac{\partial \Omega_{\rm MF}}{\partial
      \sigma} =
  \frac{\partial \Omega_{\rm MF}}{\partial \Phi} = \frac{\partial
    \Omega_{\rm MF}}{\partial \Phi^*} =0\ ,
\label{EoMMF}
\end {equation}

We will take the values $m_\pi=138$ MeV, $m_\sigma=500$
MeV, and  $f_\pi=93$ MeV in our numerical computation. The constituent quark mass in 
vacuum $m_{q}^{0}=310$ MeV fixes the value of Yukawa coupling $g=3.3$.


\section{The Proximity of the TCP to the CEP and The Phase Structure }
\label{sec:TCPCEP} 
 
The presence of CEP in the $\mu$ and T plane of the phase diagram for the real life
two flavor QCD, implies the existence of a tricritical
point (TCP) for the corresponding two massless quark flavor QCD  
because the chiral phase transition on the temperature axis turns second order 
at zero $\mu$ in the chiral limit. The second order O(4) line of critical points, 
starting from $\mu$=0 and finite T, ends at the TCP in the $\mu$ and T plane 
where it happens to meet the first order transition line originating from the chemical 
potential axis at the lowest temperature. The two flavour QMVT and PQMVT models, where 
the effect of fermionic vacuum fluctuation has been incorporated in the effective potential
of QM and PQM models, are effective QCD-like models. Hence one must find a TCP in the $\mu$ and 
T plane of the phase diagram computed in the chiral limit of zero pion mass in these models.
The present work starts with the computation of phase diagram and the location of the CEP in
the $\mu$  and T plane of all the four models QMVT,PQMVT,QM and PQM for the real life 
explicit chiral symmetry breaking with the experimental value of pion mass. Next, we locate the TCP 
in our calculation and quantify its proximity to the CEP in the phase diagram. The presence or absence 
of TCP  in the phase diagram of a model calculation and its distance from CEP, influences the nature of 
critical fluctuations around CEP.

\begin{figure*}[!tbp]
\subfigure[]{
\label{fig:pdg:a} 
\begin{minipage}[]{0.45\linewidth}
\centering \includegraphics[height=8.6cm,width=5.6cm,angle=-90]{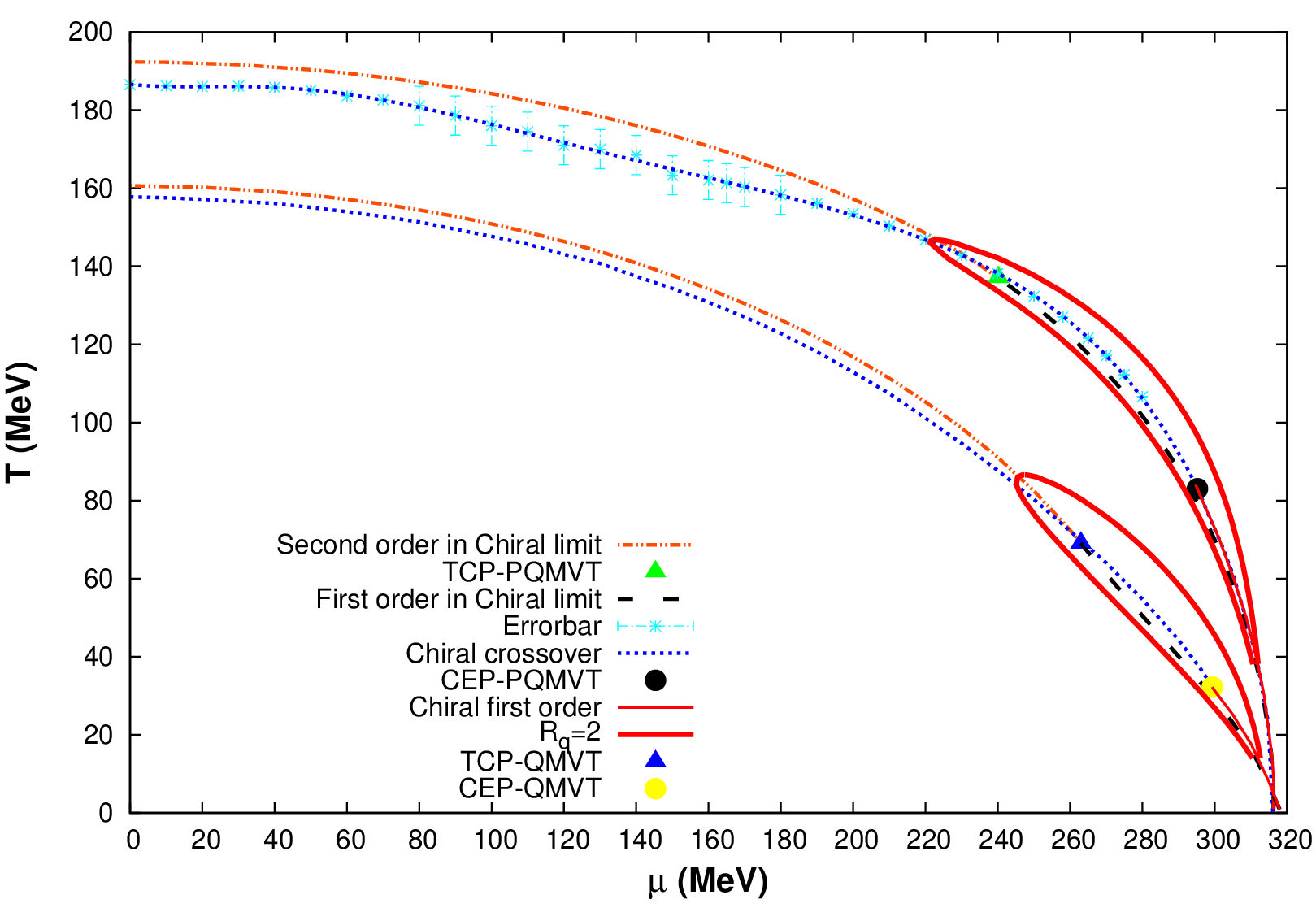}
\end{minipage}}%
\hspace{-0.03in}
\subfigure[]{
\label{fig:pdg:b} 
\begin{minipage}[]{0.45\linewidth}
\centering \includegraphics[height=8.6cm,width=5.6cm,angle=-90]{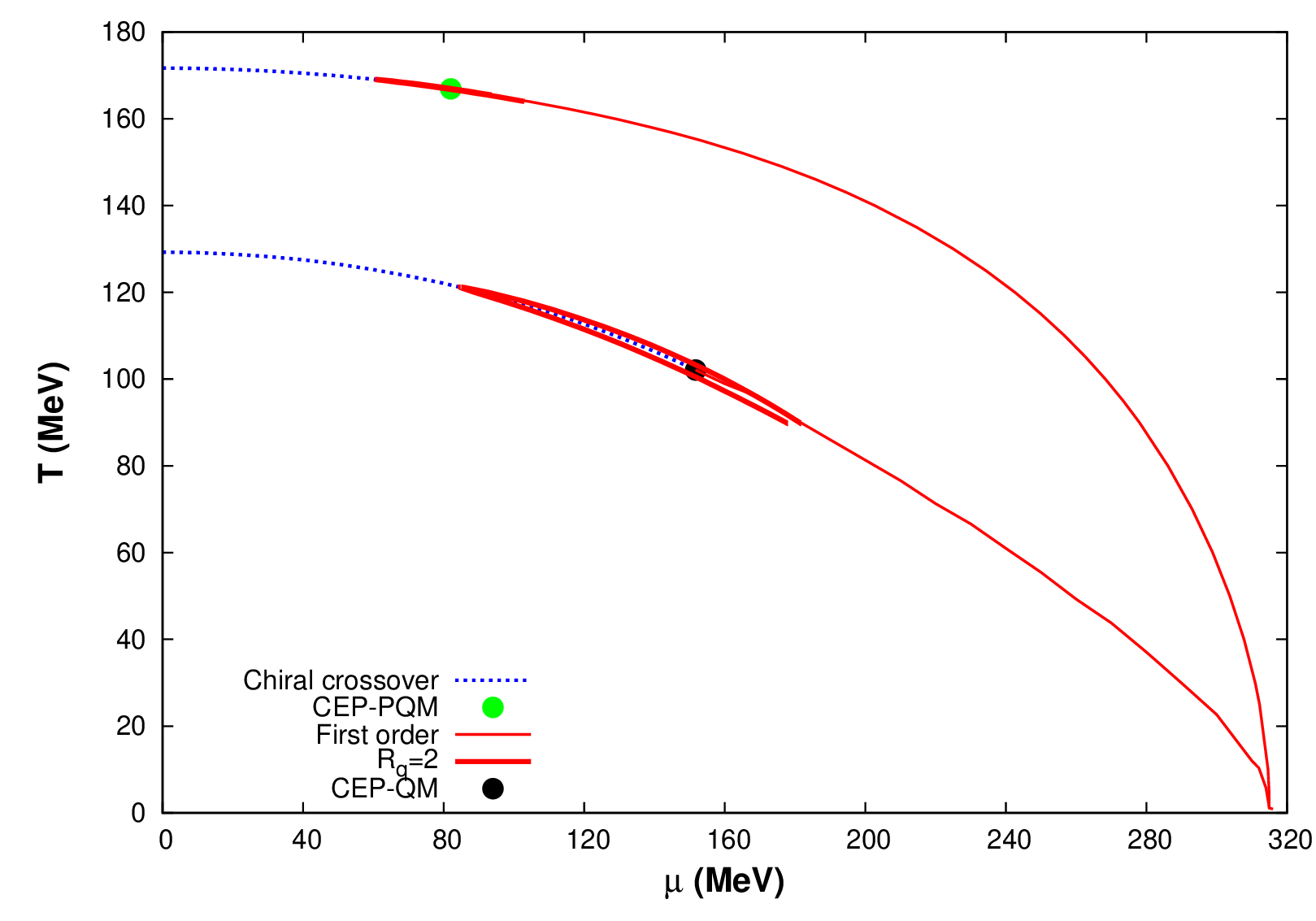}
\end{minipage}}
\caption{(a) For calculations with experimental pion mass, 
solid lines representing the first order chiral phase transition merge with the dotted
lines (blue in color) for the chiral crossover at the CEP  which is denoted by the filled circle.
The thick solid lines around CEP are the contours of constant normalized quark number 
susceptibility $R_{q}$=2. For calculations with zero pion masss, dash lines represent 
the first order phase transition in the chiral limit of QMVT and PQMVT models 
while dash dot lines represent the second order transition and the filled
triangle is the location of TCP where these two lines merge into each other. 
Lower part of the Fig. shows the QMVT model results while upper part shows the PQMVT results.
(b) Lower part of the Fig. shows the QM model results while upper part shows the PQM results.
The line types represent the same thing as in Fig.a for calculations with experimental pion mass.}
\label{fig:pdg} 
\end{figure*}

				 The results for  QMVT and PQMVT model calculations with real life pion mass,
have been presented in Fig.\ref{fig:pdg:a} while Fig.\ref{fig:pdg:b} presents the corresponding 
results for the QM and PQM model calculations. The locations of the TCP in the $\mu$ and T plane of the phase
diagrams computed with zero pion mass in the QMVT and PQMVT models, have also been shown 
in Fig.\ref{fig:pdg:a}. The TCP does not exist in the phase diagram of QM and PQM models
in the chiral limit of zero pion mass because the phase transition, on the temperature 
axis at $\mu=0$, has been found to be of first order. For calculations with experimental pion mass, 
solid lines representing the first order chiral phase transition in Fig.\ref{fig:pdg} merge with the dotted
lines (blue in color) for the chiral crossover at the CEP (denoted by filled circle). The $\pm5$ MeV error bars
(in a range $\mu=100$ to $\mu=160$ MeV) on the dotted line in the upper part of Fig.\ref{fig:pdg:a}, signify
the ambiguity of pseudo-critical temperature determination for the chiral crossover transition in the PQMVT model
(see\cite{Vivek:12} for details) calculations. The thick solid lines around  CEP are the contours of constant
ratio ($R_{q}$=2) of quark number susceptibility obtained in a model calculation to the value of quark number
susceptibility for a free quark gas. Since quark number susceptibility diverges at the CEP, such contours 
signify the extent of critical fluctuations around CEP. The CEP in the QMVT model is located at 
$\mu_{CEP}$=299.35 MeV and $T_{CEP}$=32.24 MeV as shown by the filled circle in the lower part of 
Fig.\ref{fig:pdg:a}. It shifts to the higher value on the temperature axis at $T_{CEP}$=83.0 MeV 
and $\mu_{CEP}$=295.217 MeV in PQMVT model due to the influence of Polyakov loop potential. 

				The filled circle in the lower part of Fig.\ref{fig:pdg:b} locates the CEP in QM model at
$T_{CEP}$=102.09 MeV and $\mu_{CEP}$=151.7 MeV and again in the influence of Polyakov loop
potential, the CEP in PQM model shifts considerably towards the temperature axis at $T_{CEP}$=166.88 MeV
and $\mu_{CEP}$=81.02 MeV in the upper part of the Fig.\ref{fig:pdg:b}. If we compare the location
of CEP in QM and PQM models as shown in Fig.\ref{fig:pdg:b} to the location of CEP in QMVT and
PQMVT models in Fig.\ref{fig:pdg:a}, we find a considerably significant shift of CEP to large
chemical potential and small temperature values for QMVT and PQMVT models due to the robust influence
of fermionic vacuum term inclusion in the effective potential. Another important thing worth noticing 
for the PQMVT model phase diagram in the upper part of Fig.\ref{fig:pdg:a} is the highest value of 
temperature for the chiral crossover transition occurring on the temperature axis at $\mu=0$. 
The combined effect of Polyakov loop potential and fermionic vacuum term is responsible for this.
If one compares the phase diagram in the lower part of Fig.\ref{fig:pdg:b} for QM model with the phase 
diagram in the lower part of Fig.\ref{fig:pdg:a} for QMVT model, one immediately notices that the 
the chiral crossover transition at $\mu=0$ occurs at a higher temperature value only due to the influence of
fermionic vacuum fluctuation.
These results are the  extension of our recently reported work\cite{Vivek:12} and  facilitate the details 
of model comparison for the two quark flavour case. Further these results are also in qualitative 
agreement with the recent results of Schaefer et. al.\cite{Schaef:12} for the 2+1 flavour case.

				For calculations with zero pion masss, dash lines represent the first order phase transition in 
the chiral limit of QMVT and PQMVT models in Fig.\ref{fig:pdg:a} while dash dot lines represent
the second order transition and the filled triangle is the location of TCP where these two lines
merge into each other. In the upper part of the  Fig.\ref{fig:pdg:a}, the filled triangle locates the 
presence of tricritical point (TCP) at $T_{t}$=137.09 MeV and $\mu_{t}$=240.14 MeV for PQMVT model 
calculation. In order to quantify the proximity of TCP to the CEP, we have plotted the constant 
normalized quark-number susceptibility ($R_{q}$=2) contour around CEP. It is seen
on the phase diagram that the range and extension of this contour is quite large in both the directions;chemical potential
as well as the temperature. The second cumulant of the net quark number fluctuations on this contour is double to that of 
the free quark gas value and such enhancements are the signatures of CEP for the heavy-ion collision experiments. The TCP
location is quite well inside this contour on the phase diagram. It means that the shape of the critical region and nature 
of criticality around CEP, gets influenced by the presence of TCP in the corresponding chiral limit. 
In a recent NJL/PNJL model calculation by Costa et. al.\cite{Sousa}, the CEP lies closer to the chemical potential
axis but the TCP gets located on the periphery of $R_{q}$=2 contour around CEP. In the QMVT model calculation,
the tricritical point (TCP) is found at  $T_{t}$=69.06 MeV and $\mu_{t}$=263.0 MeV as denoted by filled triangle 
in the lower part of the Fig.\ref{fig:pdg:a}. Here also the TCP lies quite well inside the $R_{q}$=2 contour on 
the phase diagram.%
%
%
%
\subsection{ Susceptibility Contours and Criticality}
\label{sec:CTRCRIT}

In order to locate the CEP in heavy-ion collision experiments, one requires the quantification of
criticality around CEP. The crossover transition is marked by a peak in the
quark number susceptibility which becomes sharper and higher as one approaches the CEP in the phase diagram
from the crossover side and finally the peak diverges at CEP. Hence the quark number susceptibilities 
and scalar susceptibilities will be significantly enhanced in a region around the CEP in the $\mu$ and T plane in 
comparison to their respective values for the free quark gas. Thus the contour regions of properly
normalized constant quark number susceptibilities and scalar susceptibilities, can be taken as the measure of
criticality around CEP. The ratio of quark-number susceptibility $\chi_q$ normalized to the free
susceptibility $\chi_q^{\text{free}}$ is written as:
\begin{equation}
  R_{q}= \frac{\chi_q}{\chi_q^{\text{free}}}
\end{equation}
The expression of quark number susceptibility is obtained as 
\begin{eqnarray}
\chi_{q}= -\frac{\partial^2 \Omega_{\rm MF}}{\partial \mu^2}\, \\ 
\label{eq:qsuscept}
\lim_{m_q\to 0}\chi_q (T, \mu)\!\!\! &=&\!\!\! \frac{\nu_q} 6 \left[ T^2 +
  \frac{3\mu^2}{\pi^2}\right]\equiv \chi_q^{\rm free}\,\\
 \nu_q= 2 N_c N_f=12
\end{eqnarray}

The first and second partial derivatives of $\sigma$, $\Phi$ and $\Phi^*$ fields with respect to 
chemical potential contribute in the double derivatives of $\Omega(\sigma)$,
$\cal U_{\text{log}}$ and $\Omega_{\mathrm{q\bar{q}}}^{\rm T}$
with respect to chemical potential as given in the appendix A of Ref.\cite{Vivek:12}.

\begin{figure*}[!tbp]
\subfigure[]{
\label{fig:QMBctr:a} 
\begin{minipage}[]{0.45\linewidth}
\centering \includegraphics[height=8.6cm,width=5.6cm,angle=-90]{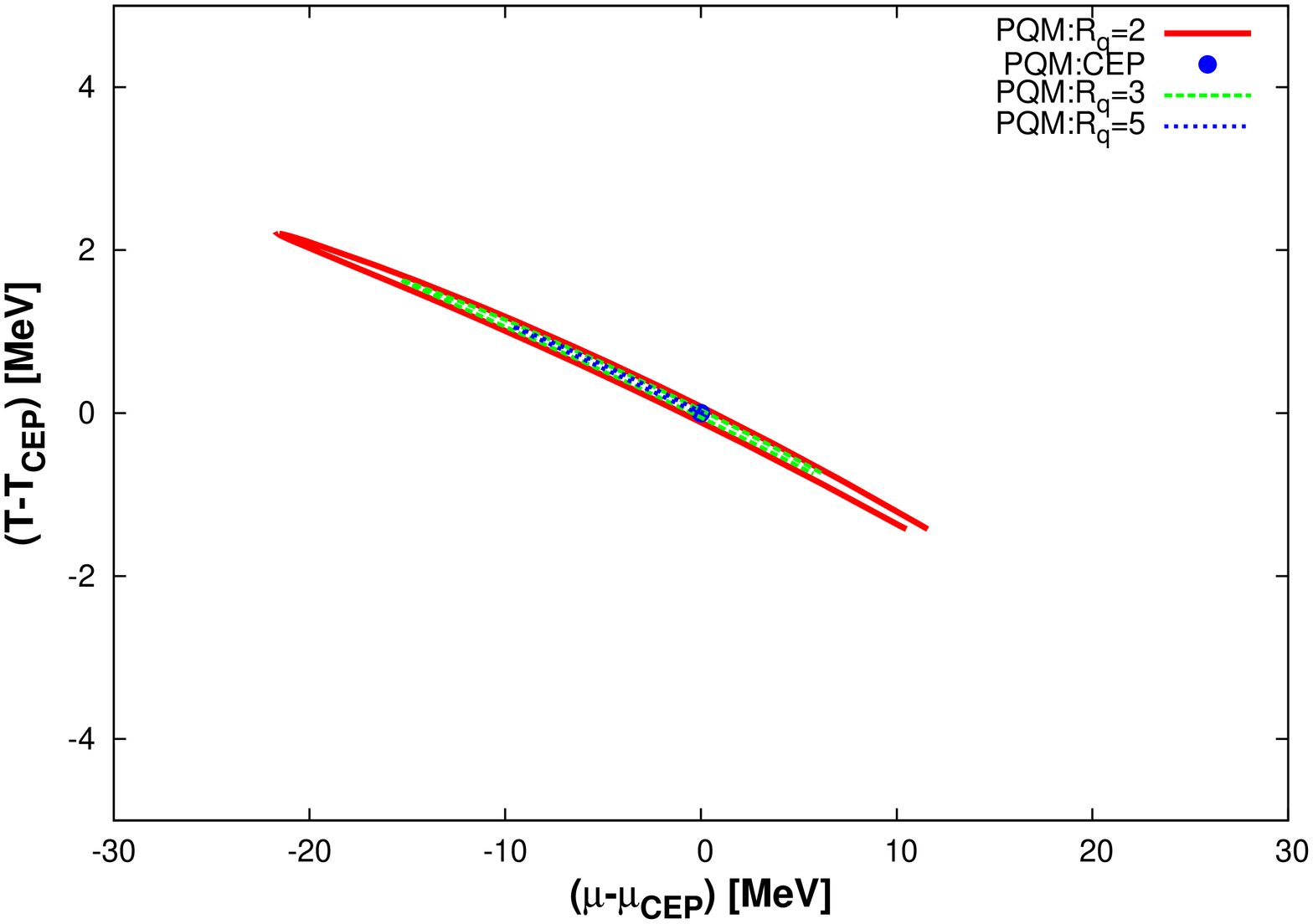}
\end{minipage}}%
\hspace{-0.03in}
\subfigure[]{
\label{fig:QMBctr:b} 
\begin{minipage}[]{0.45\linewidth}
\centering \includegraphics[height=8.6cm,width=5.6cm,angle=-90]{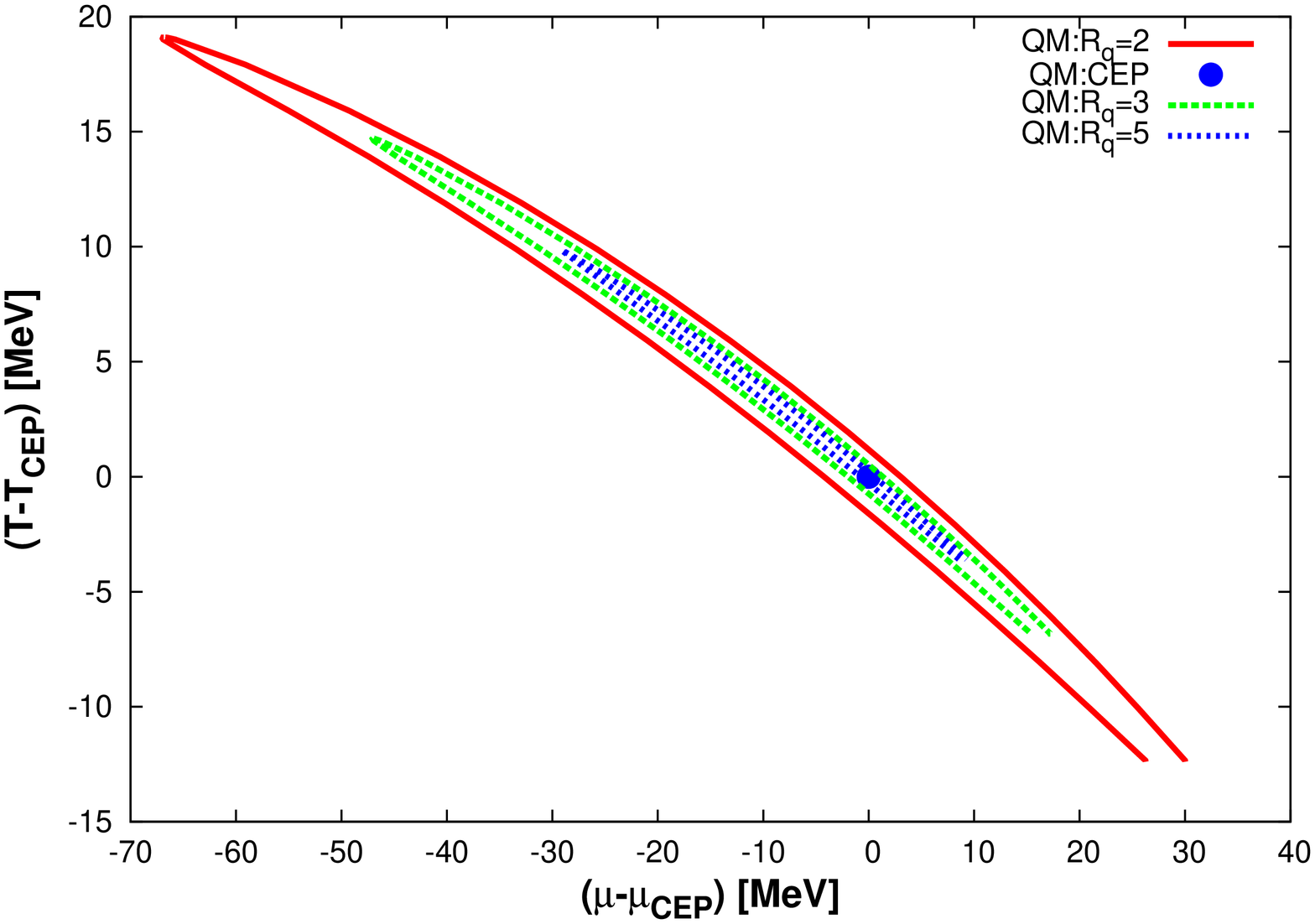}
\end{minipage}}
\caption{(a) The contours of three different values for the constant 
ratios $R_{q}=2,3$ and 5 of quark number susceptibility to the 
quark susceptibility for the free quark gas, are plotted  in the PQM 
model calculations.(b) Similar contours plotted in the QM model
calculations.}
\label{fig:QMBctr} 
\end{figure*}

				Contours with three different values for the ratios $R_q$, have been plotted in Fig.\ref{fig:QMBctr}
in the $\mu$ and T plane relative to the CEP. If we compare the contours in Fig.\ref{fig:QMBctr:a} depicting
the PQM model results to the contours in Fig.\ref{fig:QMBctr:b} showing the pure QM model results, we 
conclude that the presence of Polyakov loop potential, compresses
the critical region particularly in the T direction similar to findings of Schaefer et. al.\cite{Schaef:12}
in their three flavour calculation. The compression of critical region in the T direction is much more 
pronounced in our two quark flavour calculation as can be seen in the spread of $R_{q}=2$ contour on the 
temperature axis only in a small range of $\pm 2.5$ MeV near $T_{CEP}$. The modification in the $\mu$-direction
is quite moderate compared to the effect in the T direction. Since the chiral crossover transition becomes faster
and sharper due to the presence Polyakov loop contribution in the effective potential, the critical region in
the T direction gets significantly compressed.

\begin{figure*}[!tbp]
\subfigure[]{
\label{fig:RPQMBctr:a} 
\begin{minipage}[]{0.45\linewidth}
\centering \includegraphics[height=8.6cm,width=5.6cm,angle=-90]{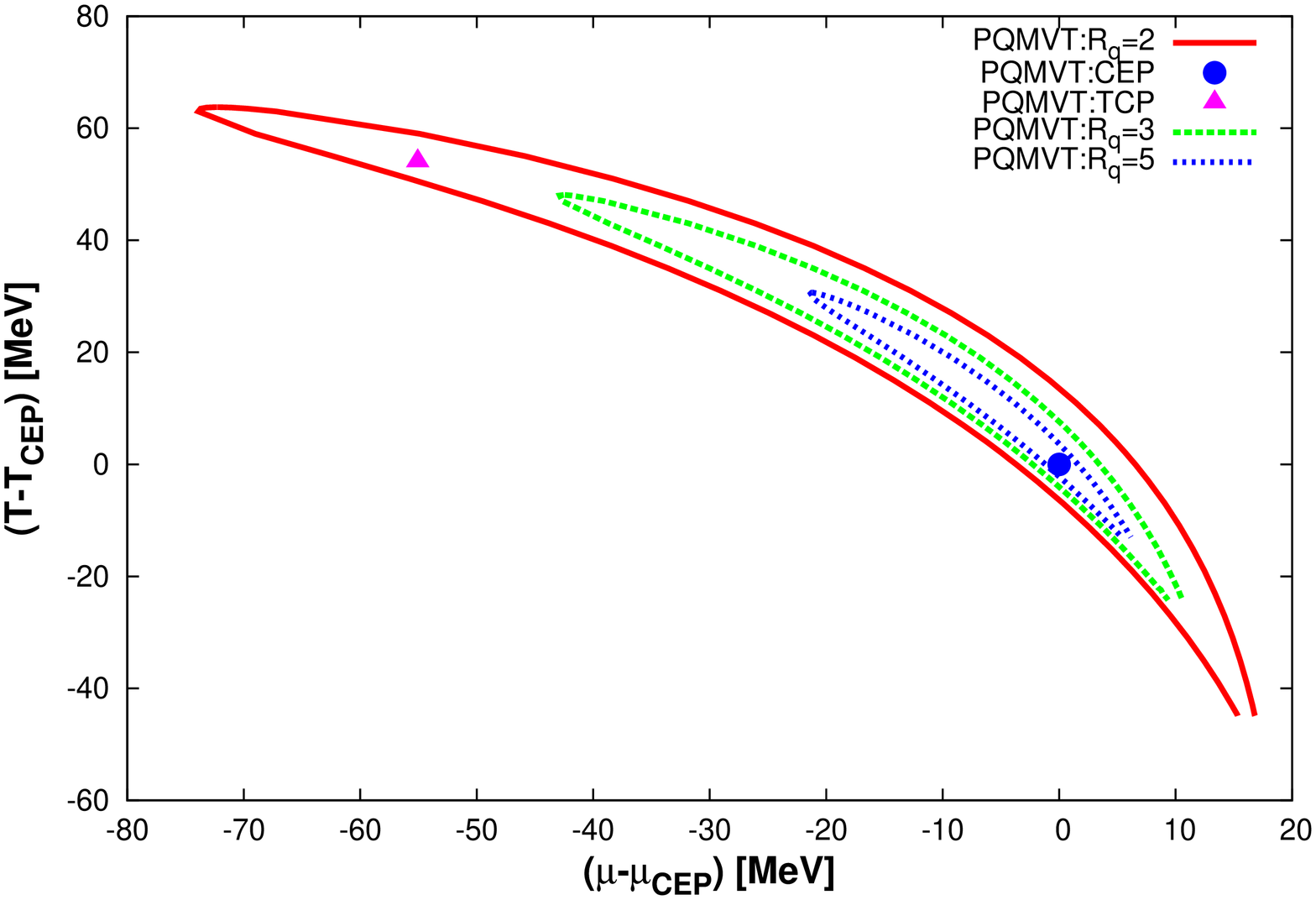}
\end{minipage}}%
\hspace{-0.03in}
\subfigure[]{
\label{fig:RPQMBctr:b} 
\begin{minipage}[]{0.45\linewidth}
\centering \includegraphics[height=8.6cm,width=5.6cm,angle=-90]{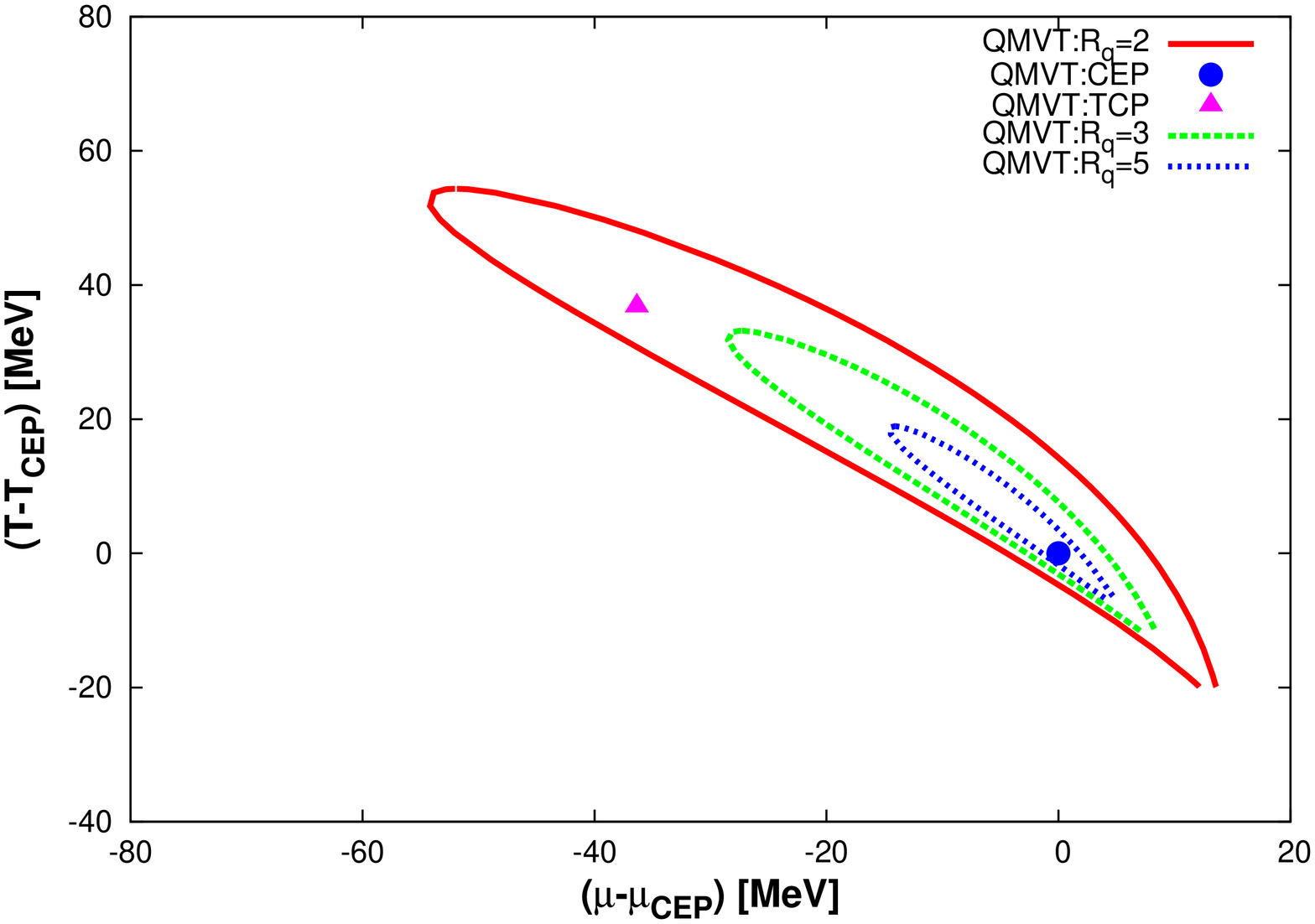}
\end{minipage}}
\caption{(a) The contours of three different values for the constant ratios 
$R_{q}=2,3$ and 5 of quark number susceptibility to the quark susceptibility 
for the free quark gas,are plotted  in the PQMVT model 
calculations.(b) Similar contours  as plotted in the QMVT model
calculations.}
\label{fig:RPQMBctr} 
\end{figure*}

		The size of the critical region is significantly influenced by the incorporation of fermionic vacuum 
fluctuations in the effective potential as shown in Fig.\ref{fig:RPQMBctr}. In the phase diagram, the size
of critical region is increased in a direction perpendicular to the crossover line due to
the influence of the fermionic fluctuations. This effect is less pronounced in Fig.\ref{fig:RPQMBctr:a} because of
the compression of critical region width due to the presence of Polyakov loop potential contribution in PQMVT model
while the QMVT model results of Fig.\ref{fig:RPQMBctr:b} obtained in the absence of Polyakov loop, show a robust 
increase in the width of the critical region. However, the extent and size of critical region in the PQMVT 
model in Fig.\ref{fig:RPQMBctr:a} is noticeably larger in both the directions $\mu$ as well as T compared 
to that of  QMVT model results as shown in Fig.\ref{fig:RPQMBctr:b}. In the presence of fermionic vacuum term,
CEP gets located at larger chemical potentials in QMVT/PQMVT models. Since the quark determinant gets modified 
mostly at moderate chemical potentials by the presence of Polyakov loop potential and further in its influence, 
the PQMVT model CEP shifts to a higher critical temperature [cf. also Fig.\ref{fig:pdg}] when compared to the 
CEP in QMVT model, we obtain an enhancement of the critical region in PQMVT model.Further the chiral crossover 
transition becomes much smoother because the phase transitions in general get washed out in the influence
of fluctuations. This leads to a critical region which is broader in  perpendicular direction to the extended 
first-order transition line. The influence of the Polyakov loop potential
becomes insignificant for smaller temperatures and larger chemical potentials, hence the size of the critical region
for $(\mu-\mu_{CEP})>0$ becomes comparable in both the models QMVT and PQMVT.
In Fig.\ref{fig:RPQMBctr}, filled circles are the position of CEP and the filled triangles, show the 
location of TCP, we observe that the TCP gets located outside $R_{q}=3$ and quite well inside the $R_{q}=2$ contour. 

			If we compare our two quark flavour results with the 2+1 flavour calculations in renormalized PQM/QM models
in Ref.\cite{Schaef:12}, we notice that in the absence of strange quarks, the effect of fermionic vacuum term, 
leads to an enhanced critical region in both the directions T as well as $\mu$ and the size of contours is 
larger in our two flavour calculation. We point out that the 2+1 flavour calculation in Ref.\cite{Schaef:12} was
done with $m_{\sigma}=400$ MeV and $T_{0}=270$ MeV while in our two flavour calculation $m_{\sigma}=500$ MeV and 
$T_{0}=208$ MeV. In general higher value of $m_{\sigma}$ pushes the CEP to higher chemical potential. In our two
quark flavour calculation, the CEP is at ($T_{CEP},\mu_{CEP}$)=(83.0,295.217) MeV and (32.24,299.35) MeV respectively
in PQMVT and QMVT model calculations while the CEP in the corresponding the 2+1 flavour model calculation of 
Ref.\cite{Schaef:12} is at ($T_{CEP},\mu_{CEP}$)=(90.0,283.0) MeV and (32,286) MeV. 

			The zero-momentum projection of the scalar propagator, encodes all fluctuations of
the order parameter and it corresponds to the scalar susceptibility $\chi_\sigma$.
The relation of the scalar susceptibility to the order parameter is obtained
as\cite{Hatta,Fujii,Ohtani,Schaefer:2006ds}. 

\begin{equation}
\label{eq:chisig}
\chi_\sigma = -\frac{\partial^2 \Omega_{\rm MF}}{\partial h^2}\
\end{equation}

The most rapid change of the chiral order parameter, should be coincident with the maximum
in the temperature or quark chemical potential variation of $\chi_\sigma$. The relation of scalar
susceptibility to the sigma mass via $\chi_\sigma \sim m_\sigma^{-2}$ can be easily verified.
The normalized scalar susceptibility is written as\cite{Schaefer:2006ds}
\begin{equation}
R_s (T,\mu)= \frac{\chi_\sigma(T,\mu)}{\chi_\sigma(0,0)}
\end{equation}
\begin{figure*}[!tbp]
\subfigure[]{
\label{fig:SPaQMctr:a} 
\begin{minipage}[]{0.45\linewidth}
\centering \includegraphics[height=8.6cm,width=5.6cm,angle=-90]{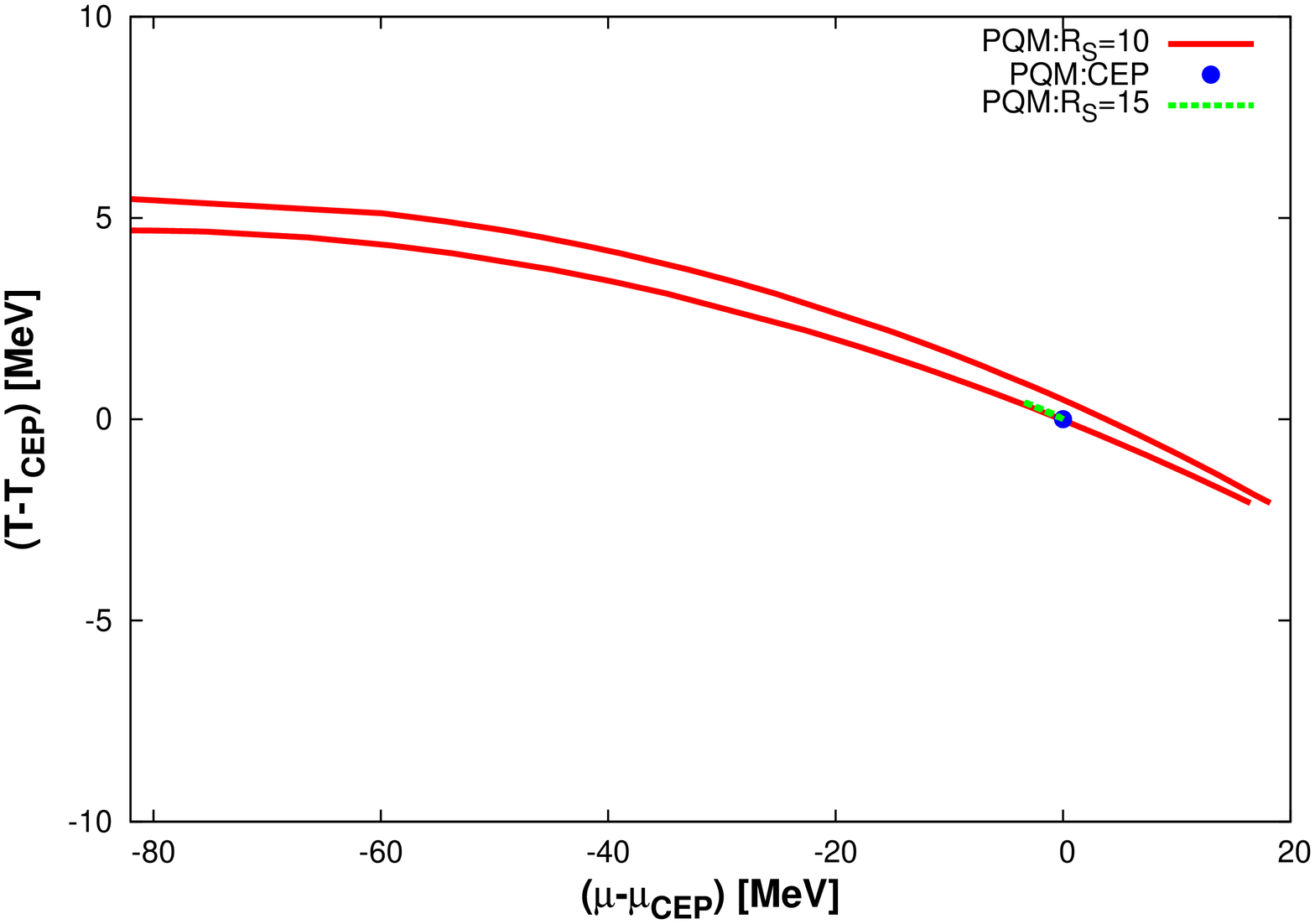}
\end{minipage}}%
\hspace{-0.03in}
\subfigure[]{
\label{fig:SPaQMctr:b} 
\begin{minipage}[]{0.45\linewidth}
\centering \includegraphics[height=8.6cm,width=5.6cm,angle=-90]{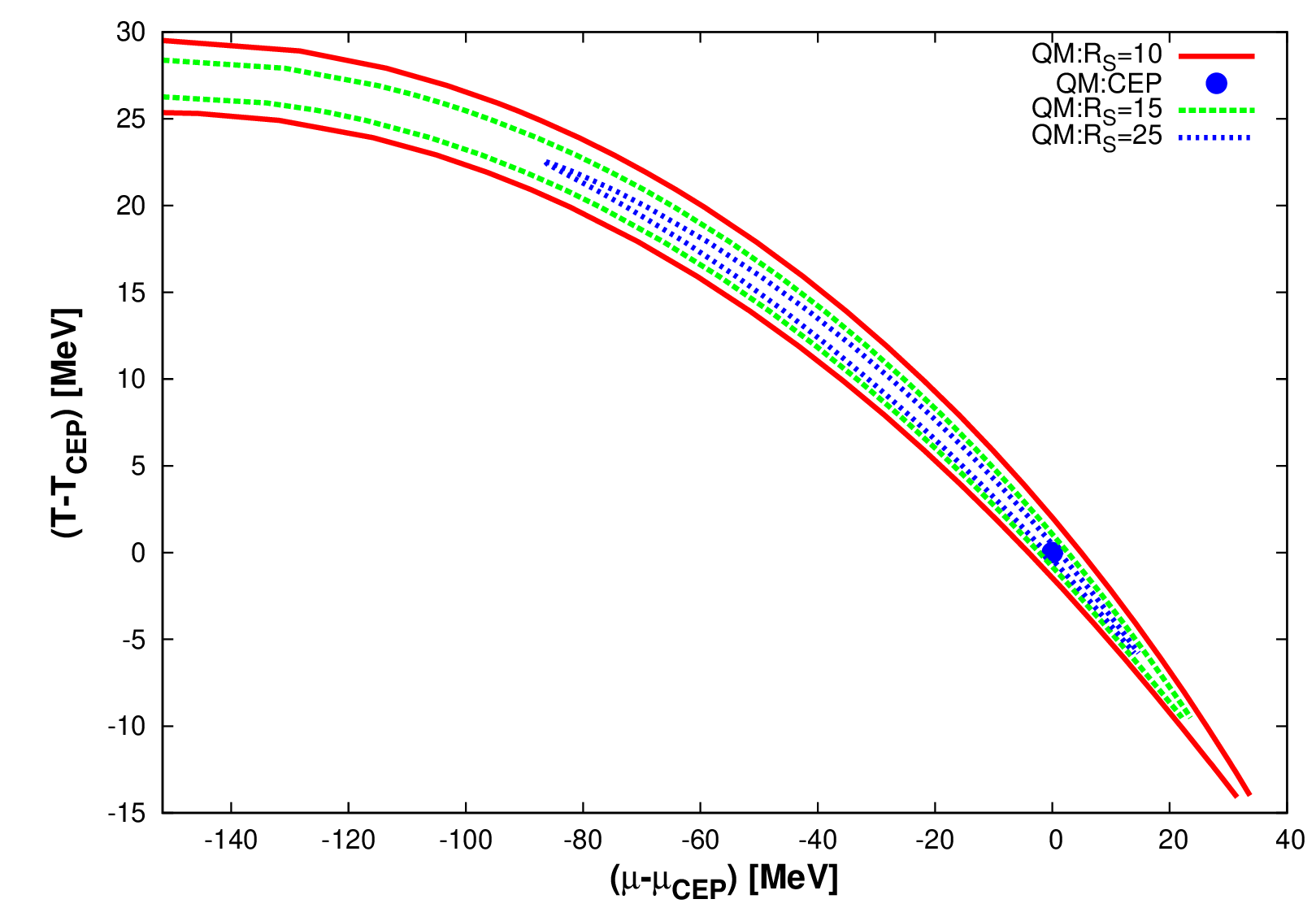}
\end{minipage}}
\caption{(a) The contours of  different values for the constant ratios
$R_{S}=10$ and 15 of T and $\mu$ dependent scalar susceptibility to the scalar 
susceptibility at T=0 and $\mu$=0 MeV, are plotted  in the PQM model 
calculations.(b) Similar contours for the constant ratios $R_{S}=10,15$ and 25
are plotted in the QM model calculations. }
\label{fig:SPaQMctr} 
\end{figure*}

In Fig.~\ref{fig:SPaQMctr}, the contours have been plotted for
three values of fixed ratios $R_{s}$ around the CEP in the PQM and QM models. The $R_{s}=10$ contour
in Fig.~\ref{fig:SPaQMctr:a} is compressed in the T direction and its extension in 
$\mu$ direction is also reduced in comparison to the pure QM model contours in Fig.~\ref{fig:SPaQMctr:b}.
This is due to the quite fast and rapid temperature or chemical potential variation of $\sigma$ meson mass
$m_{\sigma}$ on account of faster and sharper change of order parameter for chiral crossover in the presence of 
Polyakov loop potential in the calculations. We do not find contour for $R_{s}=25$ in Fig.~\ref{fig:SPaQMctr:a}
because the minimum value of $\sigma$ meson mass does not fall below 100 MeV, though the value of $m_{\sigma}$
falls very rapidly and sharply from 500 MeV to 128 MeV giving rise to a very thin and small contour region
even for $R_{s}=15$. In the QM model calculations, we get all the contour regions for $R_{s}=10,15$ and 25 with well
defined size because the $m_{\sigma}$ variation is smoother and slower in comparison to the corresponding PQM model
results and further the minimum in the $m_{\sigma}$ variation approaches almost zero value in QM model. 
The chiral crossover transition on the temperature axis at $\mu=0$ MeV in the QM and PQM models, is quite sharp and 
fast because it emerges from the background of first order chiral transition at $\mu=0$ MeV  in the corresponding chiral
limit of zero pion mass and we do not find the existence of TCP in QM/PQM models. As a consequence, we
do not find the closure of $R_{s}=10$ contour on the temperature axis at $\mu=0$ MeV.

\begin{figure*}[!tbp]
\subfigure[]{
\label{fig:SRPaQMctr:a} 
\begin{minipage}[]{0.45\linewidth}
\centering \includegraphics[height=8.6cm,width=5.6cm,angle=-90]{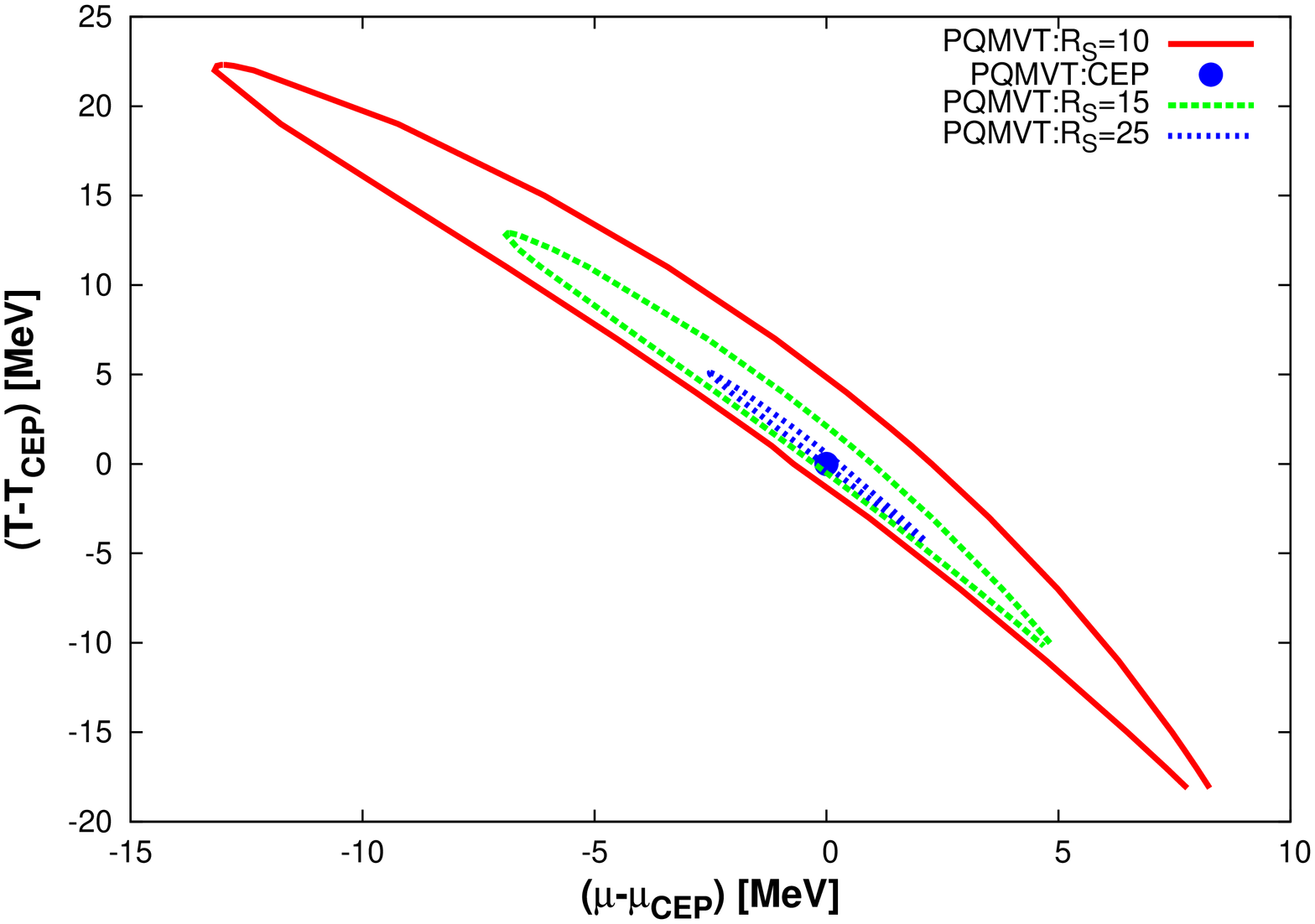}
\end{minipage}}%
\hspace{-0.03in}
\subfigure[]{
\label{fig:SRPaQMctr:b} 
\begin{minipage}[]{0.45\linewidth}
\centering \includegraphics[height=8.6cm,width=5.6cm,angle=-90]{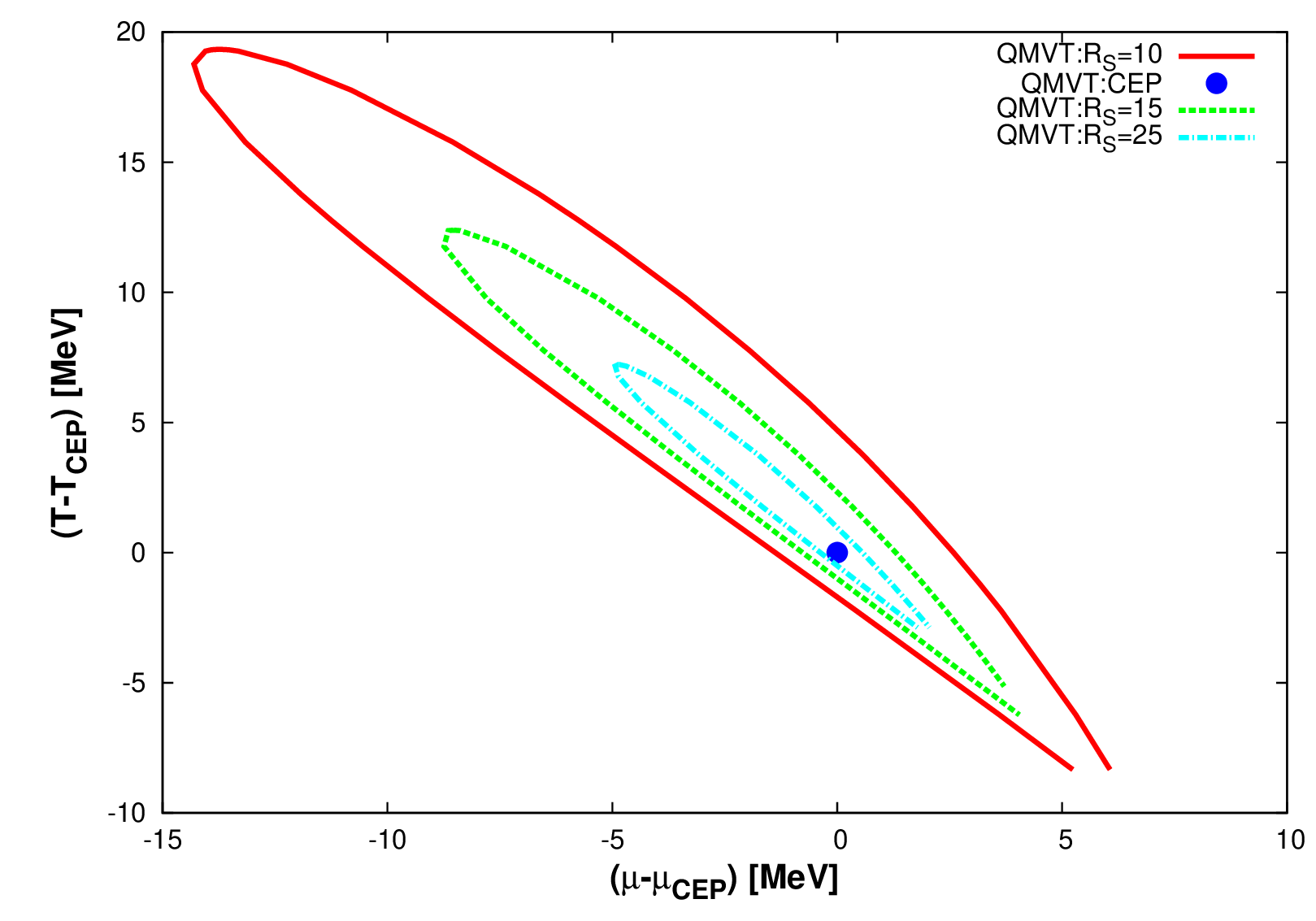}
\end{minipage}}
\caption{(a) The contours of three different values for the constant ratios 
$R_{S}=10,15$ and 25 of T and $\mu$ dependent scalar susceptibility to the scalar
susceptibility at T=0 and $\mu$=0 MeV,are plotted  in the PQMVT model 
calculations.(b)Similar contours for the constant ratios $R_{S}=10,15$ and 25
are plotted in the QMVT model calculations.}
\label{fig:SRPaQMctr} 
\end{figure*}

			We obtain quite well defined and closed contour regions for $R_{s}=10,15$ and 25 in Fig.~\ref{fig:SRPaQMctr} which 
again become broader in the  direction perpendicular to the crossover line due to the presence of fermionic  vacuum fluctuations
in QMVT and PQMVT model calculations. For scalar susceptibility also, the critical region gets elongated  in the phase diagram and
$\chi_\sigma$ is enhanced in the direction parallel to the first-order transition line. Here also the presence of Polyakov loop 
potential in the PQMVT model, leads to the compression in the width of critical region around CEP as shown in  Fig.\ref{fig:SRPaQMctr:a}.
The fermionic vacuum fluctuations, make the chiral crossover transition very smooth while the Polyakov loop potential makes 
it sharper and faster and these opposite effects give a typical shape to the quark number susceptibility contours 
in Fig.\ref{fig:RPQMBctr:a} in the  PQMVT model. Similar effects can be seen in the scalar susceptibility contours 
in Fig.\ref{fig:SRPaQMctr:a}. In the influence of fermionic vacuum fluctuations only, the $\chi_\sigma$ contours  
in Fig.\ref{fig:SRPaQMctr:b} in the pure QMVT model are broader and rounded. 

	For the detail understanding and analysis of the criticality around the CEP, 
we will be studying the critical exponents of the susceptibilities at 
the critical point in the next section.

\subsection{ Critical Exponents }
\label{subsec:Expncrit} 

	The crossover transition is marked by a peak in the quark number susceptibility which diverges 
as one approaches the CEP from the crossover side in the phase diagram. This divergence is
governed by a power law within the critical region. The corresponding critical exponents 
depend on the route through which the singularity (CEP) is approached in the $\mu$ and 
T plane \cite{Griffiths1970}. This path dependence decides the shape of the critical region.
In the mean-field approximation, the quark number susceptibility scales with an exponent $\gamma_q=1$
for a path asymptotically parallel to the first-order transition line and for any other path which is not
parallel to the first-order line,the divergence scales with the exponent $\epsilon = 2/3$. 
This larger critical exponent ($\gamma_q > \epsilon$) is one reason for the elongation of the critical region 
in a direction parallel to the first-order line as already pointed out 
in\cite{Hatta, Schaefer:2006ds}. 

\begin{figure*}[!tbp]
\subfigure[]{
\label{fig:expcrRPQM:a} 
\begin{minipage}[]{0.45\linewidth}
\centering \includegraphics[height=6.2cm,width=8.8cm,angle=0]{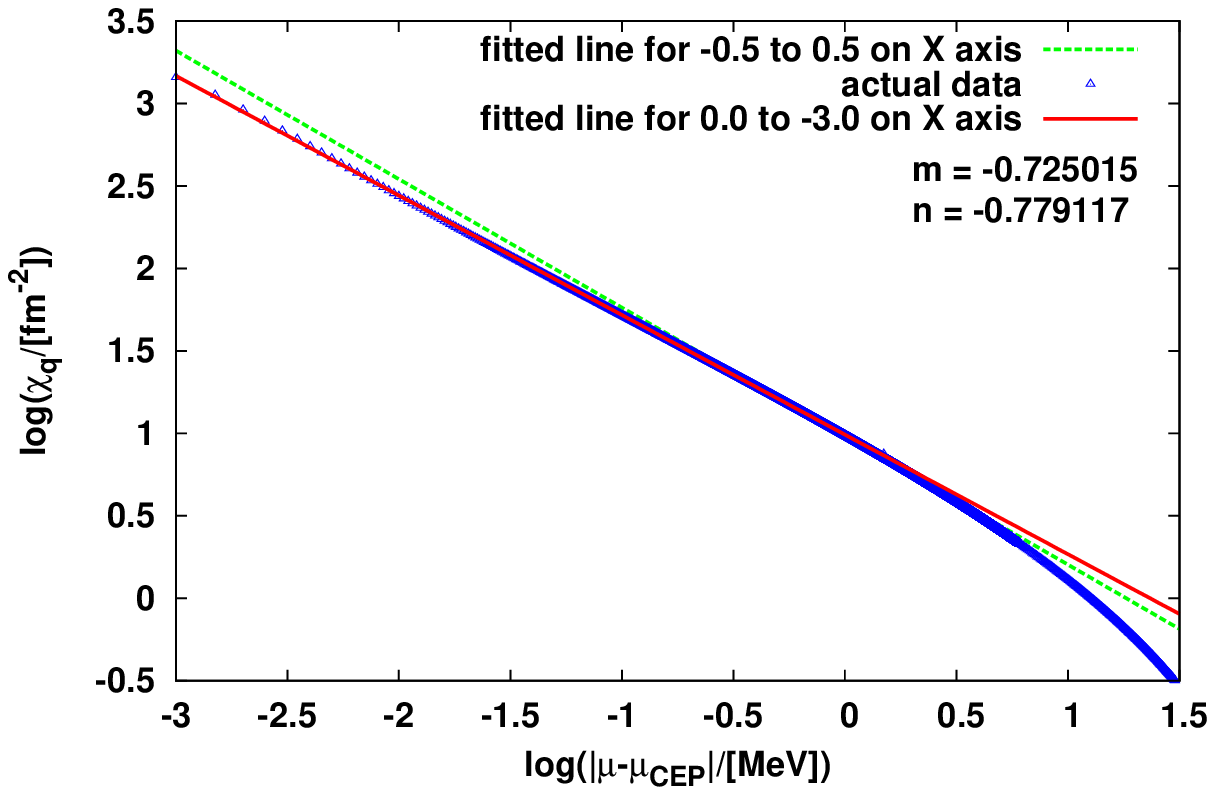}
\end{minipage}}%
\hspace{-0.001in}
\subfigure[]{
\label{fig:expcrRPQM:b} 
\begin{minipage}[]{0.45\linewidth}
\centering \includegraphics[height=6.2cm,width=8.8cm,angle=0]{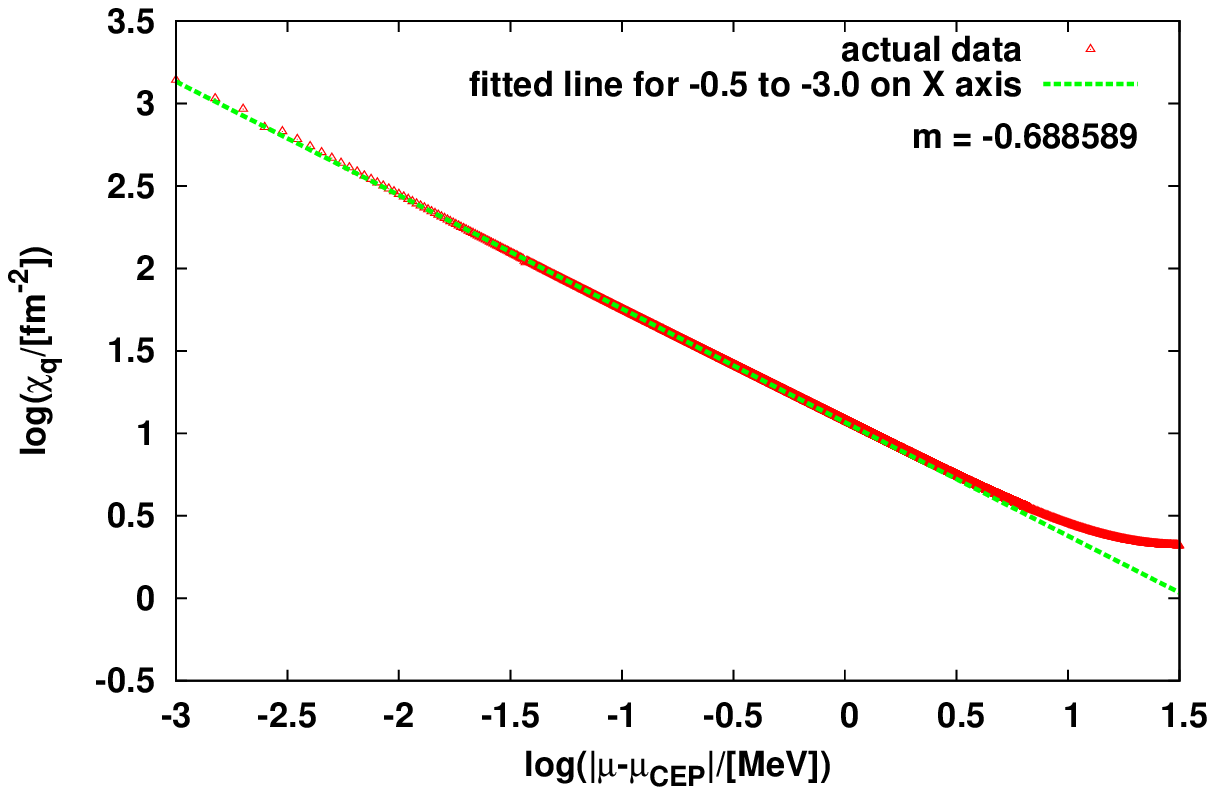}
\end{minipage}}
\caption{(a) shows the plot of the logarithm of $\chi_q$ as a function of the 
logarithm of $\mu-\mu_{CEP}$ close to the CEP in PQMVT model 
when the $\mu_{CEP}$ is approached from the lower $\mu$ side (b) shows the 
same plot as in Fig.a in PQMVT model when the $\mu_{CEP}$ is approached 
from the higher $\mu$ side.}
\label{fig:expcrRPQM} 
\end{figure*}

		In order to further investigate the nature of criticality in two flavour calculations,
we have numerically evaluated the critical exponents of the quark number 
susceptibility $\chi_q$ in QM,PQM,QMVT and PQMVT models. In these investigations,
the critical $\mu_{CEP}$ at fixed critical temperature $T_{CEP}$ is approached from 
the lower as well as higher $\mu$ sides in a path parallel to the $\mu$-axis in 
the ($T,\mu$)-plane. The calculation of the critical exponents, has been done with
the following linear logarithmic fit formula: 
\begin{equation} 
\log \chi_q = -m \log |\mu -\mu_{CEP} | + c\ ,
\end{equation} 
The slope m gives the critical exponent  $\epsilon$ and the Y axis
intercept $c$ is independent of $\mu$.
\begin{figure*}[!tbp]
\subfigure[]{
\label{fig:QMcrexp:a} 
\begin{minipage}[]{0.45\linewidth}
\centering \includegraphics[height=6.2cm,width=8.8cm,angle=0]{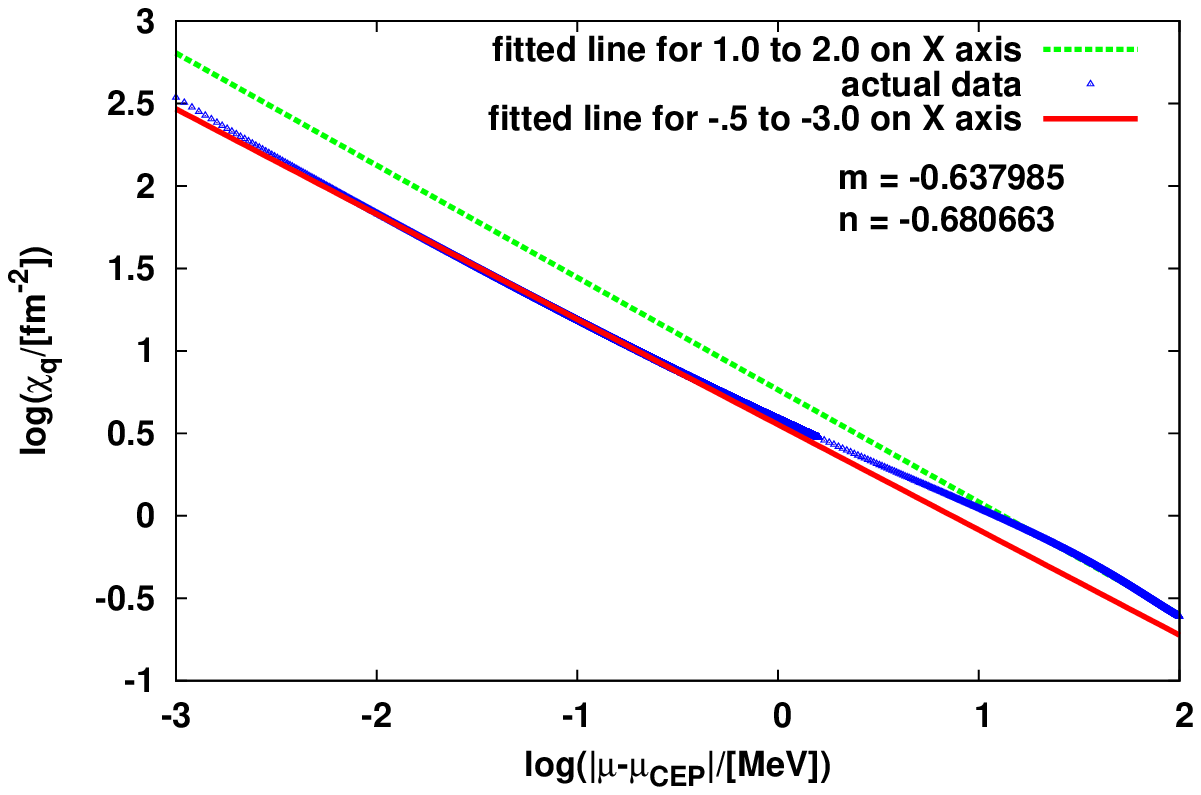}
\end{minipage}}%
\hspace{-0.001in}
\subfigure[]{
\label{fig:QMcrexp:b} 
\begin{minipage}[]{0.45\linewidth}
\centering \includegraphics[height=6.2cm,width=8.8cm,angle=0]{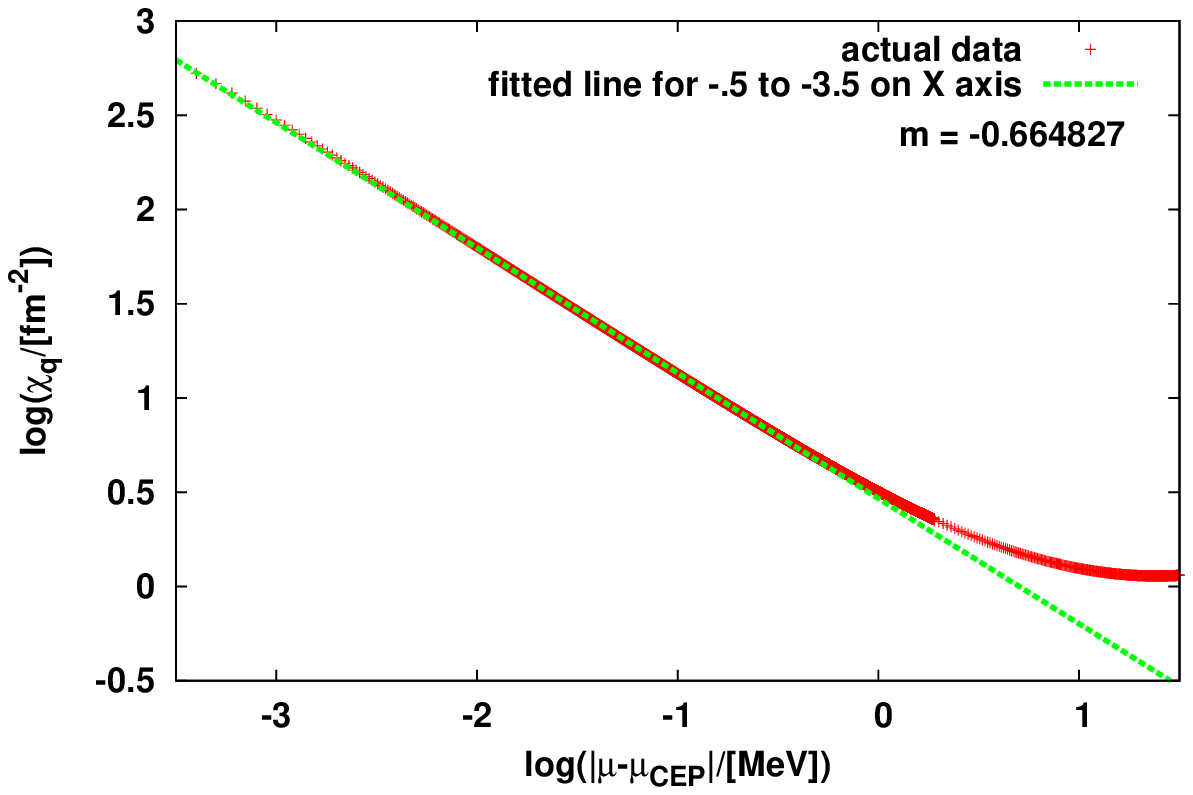}
\end{minipage}}
\caption{(a) Shows the plot of the logarithm of $\chi_q$ as a function of the 
logarithm of $\mu-\mu_{CEP}$ close to the CEP in QM model when the $\mu_{CEP}$ 
is approached from the lower $\mu$ side (b) shows the same plot as in 
Fig.a in QM model when the $\mu_{CEP}$ is approached 
from the higher $\mu$ side.}
\label{fig:QMcrexp} 
\end{figure*}
			 Fig.~\ref{fig:QMcrexp} shows the logarithm of $\chi_q$ as a function of the 
logarithm of $\mu-\mu_{CEP}$ close to the CEP in QM model. Scaling is observed over
several orders of magnitude. In Fig.~\ref{fig:QMcrexp:a}, the $\mu_{CEP}$ is approached 
from the lower $\mu$ side and we obtain a critical exponent $\epsilon=m = 0.6379\pm 0.0002$  
while the critical exponent $\epsilon=m = 0.6648\pm 0.0001$ in the result of Fig.~\ref{fig:QMcrexp:b} 
when the $\mu_{CEP}$ is approached from the higher $\mu$ side. The scaling starts around
around $\log |\mu- \mu_{CEP} | < -.5$ in both the cases. These exponents show good agreement
with the mean-field prediction $\epsilon = 2/3$. In Fig.~\ref{fig:QMcrexp:a}, the data fitted
in a range $1.0<\log |\mu- \mu_{CEP} | < 2.0$ also shows a scaling kind of linear behaviour over one order 
of magnitude with a larger slope $n=.6807\pm.0002$ which changes to $m=.6379\pm.0002$ in the range 
$-.5<\log |\mu- \mu_{CEP} | < 1.0$. When $log |\mu- \mu_{CEP} | \sim 2.0$, we are very close to $\mu=0$ on the 
temperature axis. The phase transition in the chiral limit at $\mu=0$ in QM model is first order and it
becomes crossover for the real life pion mass. In the quark mass (or the pion mass) and T plane at 
$\mu=0$, the first order transition line should change to crossover line through another second 
order critical end point as one increases the pion mass from zero to the experimental value. This
linear behaviour in a range $1.0<\log |\mu- \mu_{CEP} | < 2.0$ with larger slope may be due to the influence
of another hidden CEP in the mass and temperature plane at $\mu=0$ in the QM model. The
critical exponent values obtained in PQM model calculation have been given in Table.1, we can see that
the presence of Polyakov loop potential in QM model, does not influence the value of critical exponents.
\begin{figure*}[!tbp]
\subfigure[]{
\label{fig:RQMcrexp:a} 
\begin{minipage}[]{0.45\linewidth}
\centering \includegraphics[height=6.2cm,width=8.8cm,angle=0]{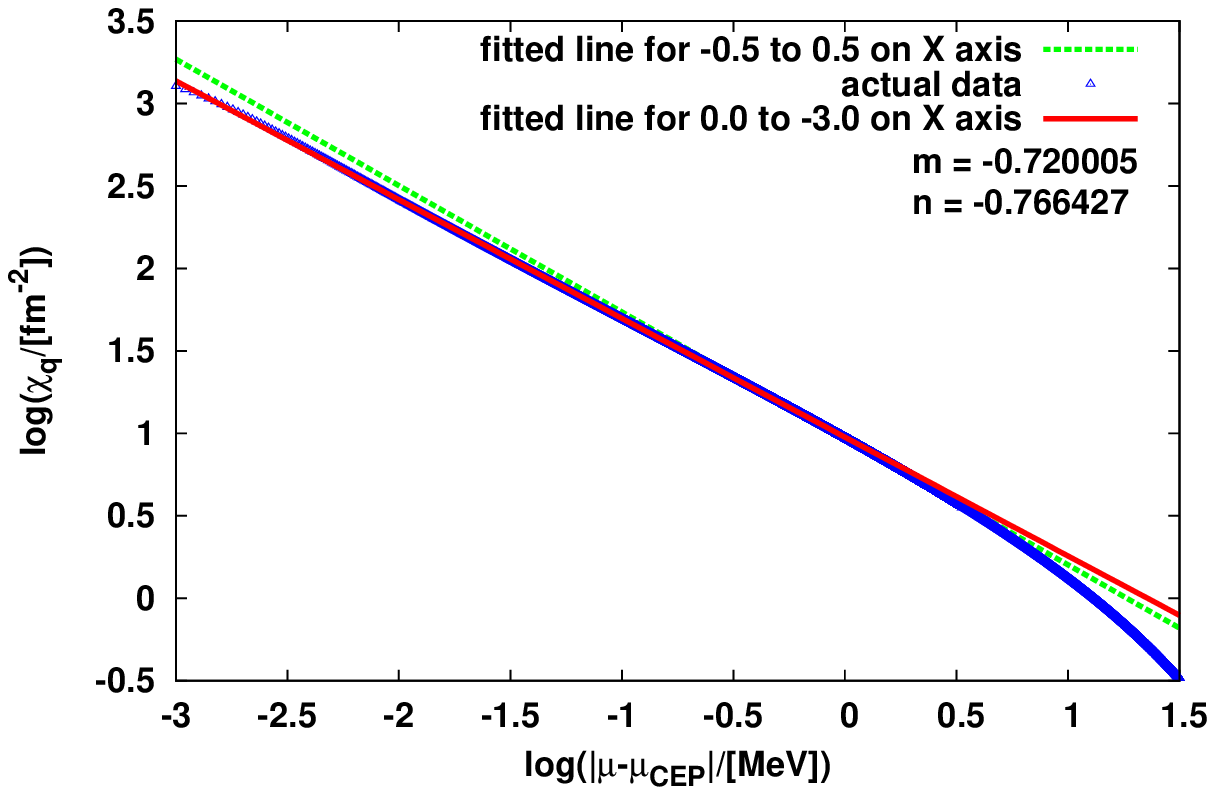}
\end{minipage}}%
\hspace{-0.001in}
\subfigure[]{
\label{fig:RQMcrexp:b} 
\begin{minipage}[]{0.45\linewidth}
\centering \includegraphics[height=6.2cm,width=8.8cm,angle=0]{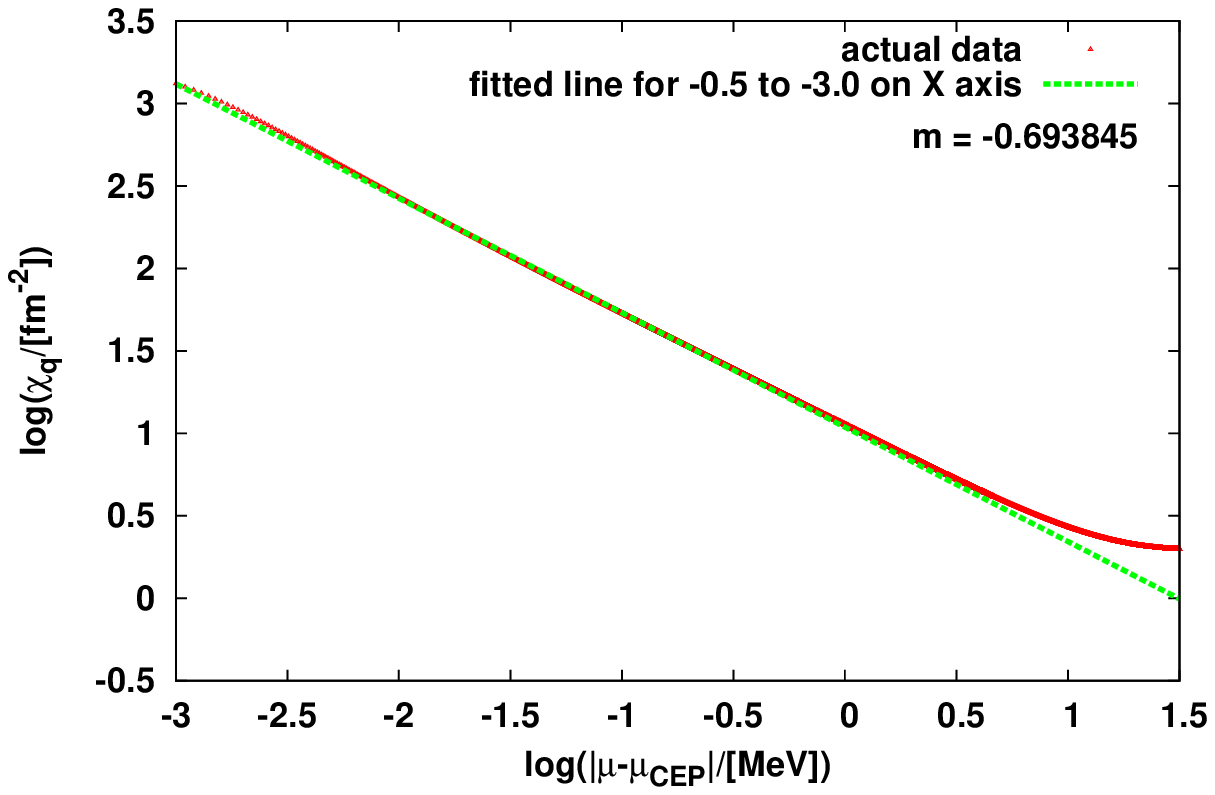}
\end{minipage}}
\caption{(a) Shows the plot of the logarithm of $\chi_q$ as a function of the 
logarithm of $\mu-\mu_{CEP}$ close to the CEP in QMVT model when the $\mu_{CEP}$
is approached from the lower $\mu$ side (b) shows the same plot as in 
Fig.a in QMVT model 
when the $\mu_{CEP}$ is approached from the higher $\mu$ side.}
\label{fig:RQMcrexp} 
\end{figure*}
 Fig.~\ref{fig:RQMcrexp} shows the plot of the logarithm of $\chi_q$ with respect to the 
logarithm of $\mu-\mu_{CEP}$ in the presence of fermionic vacuum fluctuations in QMVT model. 
In Fig.~\ref{fig:RQMcrexp:a} when the $\mu_{CEP}$ is approached 
from the lower $\mu$ side, we obtain a larger critical exponent $\epsilon=m = 0.720\pm 0.00005$ 
in comparison to the corresponding result in the QM model. Due to the influence of fermionic
vacuum fluctuation, we find the presence of TCP in the chiral limit of QMVT model and it lies
quite well within the $R_{q}=2$ contour surrounding the CEP in the phase diagram in Fig.~\ref{fig:pdg:a}. 
This larger critical exponent may be the consequence of the modification of criticality around CEP due to
the presence of TCP in its proximity. The scaling starts earlier when $\log |\mu- \mu_{CEP} | < 0.0$ and we observe
scaling over several orders of magnitude. This higher value of the critical exponent is  close to the critical exponent
calculated in Ref.\cite{Schaefer:2006ds} where the effect of quantum fluctuations in the QM model were incorporated in 
the Proper-Time Renormalization Group (PTRG) approach. The critical exponents change  in the 
range $-0.5<\log |\mu- \mu_{CEP} |<.5 $ in Ref. \cite{Schaefer:2006ds}, from .77 for the one scaling 
regime starting after $\log |\mu- \mu_{CEP} |>.5 $ to .74 for another scaling regime starting 
before $\log |\mu- \mu_{CEP} |<-.5$. Though we do not find analogous crossing behavior of the 
universality classes, the data points in our calculation show a bending trend and when we fit 
the data in a small range $-0.5<\log |\mu- \mu_{CEP} |<.5 $, we find a higher slope 
$\epsilon=m = 0.7664\pm 0.0002$ as shown in the Fig.~\ref{fig:RQMcrexp:a}. In our calculation, the
critical region of CEP is having a noticeable overlap with the critical region of TCP and this may 
be the reason of the bending trend in the data. If we approach the CEP from the higher $\mu$ side, we find smaller
critical exponent $\epsilon=m = 0.6938\pm 0.0002$ in the result of Fig.~\ref{fig:RQMcrexp:b}. In this case, 
the scaling starts around $\log |\mu- \mu_{CEP} | < -.5$ . It is pointed out that these exponents calculated
in the presence of fermionic vacuum fluctuations in the QM model are different from the mean-field
prediction $\epsilon = 2/3$. Similar results are found in Fig.~\ref{fig:expcrRPQM} which shows the 
plot of the logarithm of $\chi_q$ with respect to the logarithm of $\mu-\mu_{CEP}$ in the presence of 
fermionic vacuum fluctuations in PQMVT model. The presence of Polyakov loop compresses the width of 
critical region in PQMVT model but its effect on critical exponents is negligibly small as can be seen
in Fig.~\ref{fig:expcrRPQM:a} and Fig.~\ref{fig:expcrRPQM:b}. We point out that the recent
calculation in 2+1 flavour model in Ref.\cite{Schaef:12} does not report any modification of the
mean field critical exponents under the infulence of fermionic vacuum term. The chiral crossover transition
occurring at $\mu=0$ on the temperature axis, will be faster and sharper due to the presence of s quarks
in their 2+1 quark flavour calculation if we  compare the present two quark flavour calculation with 
that of them.

 The critical exponents calculated in all the models are summarized and tabulated in Table 1.

%
\begin{table}
\begin{tabular}{|c|c|c|}
\hline
Model &$\mu-\mu_{CEP} < 0$& $\mu-\mu_{CEP} > 0$ \\ \hline \hline
QM	  & $0.6379 \pm 0.0002$ &$0.6648 \pm 0.0001$\\

PQM		& $0.6309 \pm 0.0001$ &$0.6668 \pm 0.0001$\\

QMVT	&  $0.720\pm 0.00005$ &$0.6938 \pm 0.0002$ \\

PQMVT	& $0.725 \pm 0.0002$ &$0.6886 \pm 0.0004$\\ \hline
\end{tabular}
\caption{\label{tab:crexponets} Critical exponents of the
  quark-number susceptibility in the QM,PQM,QMVT and PQMVT models
  for two different paths parallel to the chemical potential axis 
  approaching the $\mu_{CEP}$ from the lower $\mu<\mu_{CEP}$ and 
  higher $\mu>\mu_{CEP}$side.}
\end{table}

\subsection{In medium meson masses} 
\label{subsec:Masscrit}
\begin{figure*}[!tbp]
\subfigure[]{
\label{fig:QMaRQMmass:a} 
\begin{minipage}[]{0.45\linewidth}
\centering \includegraphics[height=8.6cm,width=5.6cm,angle=-90]{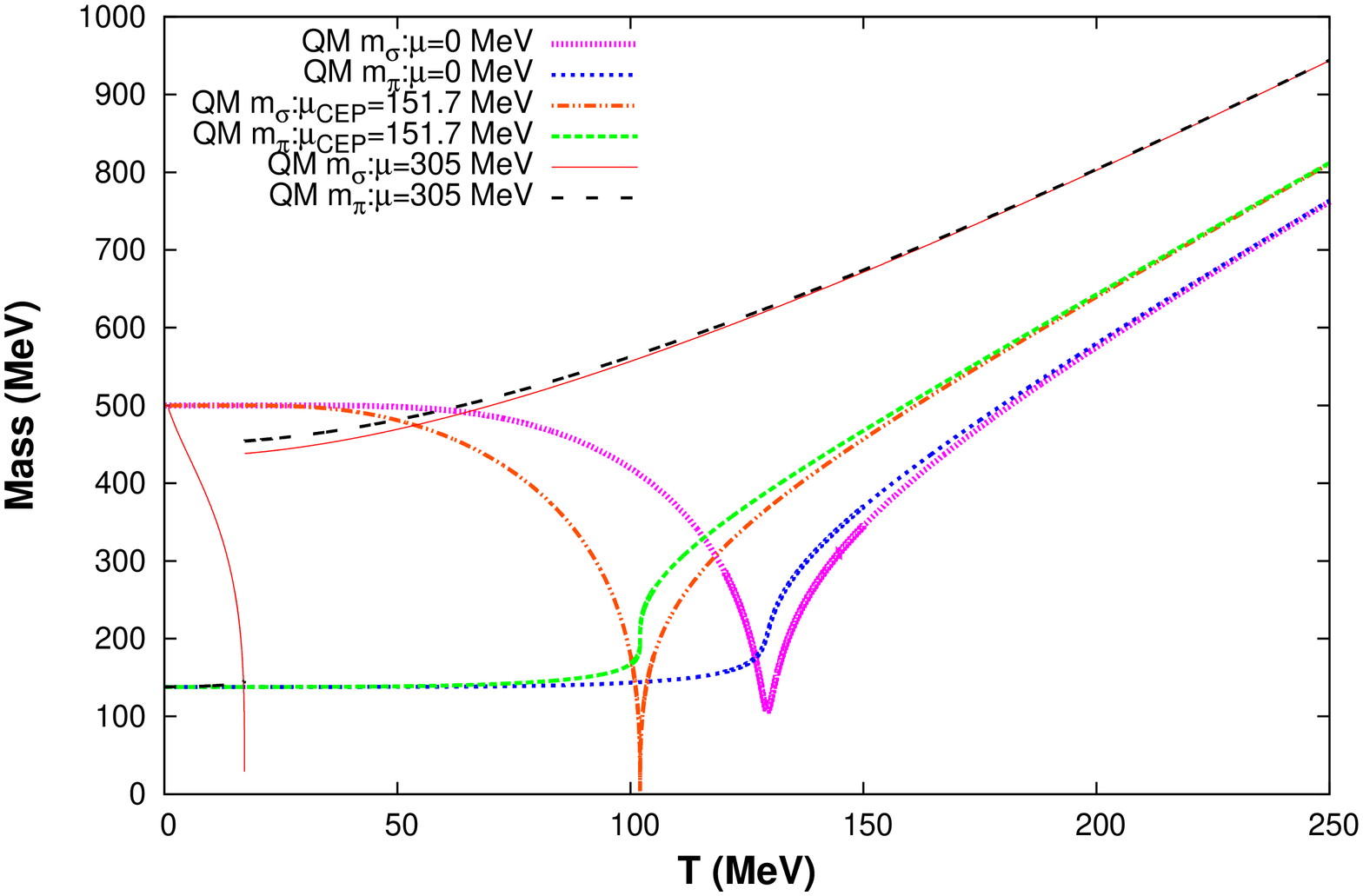}
\end{minipage}}%
\hspace{-0.03in}
\subfigure[]{
\label{fig:QMaRQMmass:b} 
\begin{minipage}[]{0.45\linewidth}
\centering \includegraphics[height=8.6cm,width=5.6cm,angle=-90]{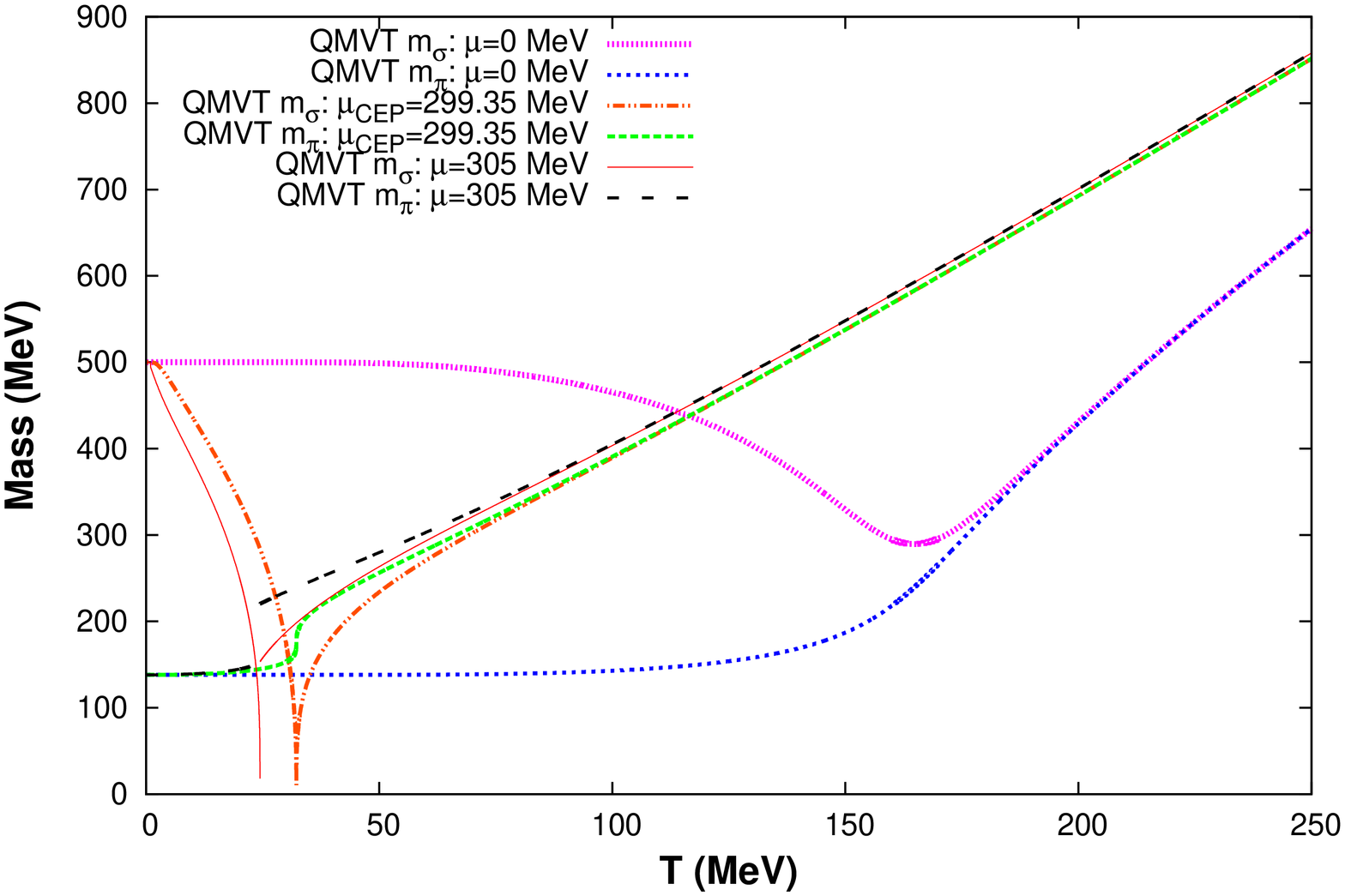}
\end{minipage}}
\caption{(a) Mass variations in the QM model are shown in this Fig.
Two dotted lines in the rightmost part of the Fig. represent the
temperature variations of $m_{\sigma}$ and $m_{\pi}$  at $\mu=0$, the dash dotted
line in the middle part represents $m_{\sigma}$ while the line with small dash 
represents $m_{\pi}$ temperature variations at $\mu_{CEP}=151.7$ and the solid
line in the leftmost part of the Fig. represents $m_{\sigma}$ while the line 
with dash represents $m_{\pi}$ temperature variations at $\mu=305$ MeV.
(b) Temperature variations of $m_{\sigma}$ and $m_{\pi}$ in the influence of 
fermionic vacuum fluctuations, have been shown for the QMVT model calculations.
Lines represent the same  mass variations as in the left panel of the Fig. In QMVT model
$\mu_{CEP}=299.35$ MeV.}
\label{fig:QMaRQMmass} 
\end{figure*}

The critical fluctuations are also encoded in the variation of 
meson masses $m_\pi(T,\mu)$ and $m_\sigma (T,\mu)$ as one passes
through the chiral symmetry restoring phase transition.
We will investigate and compare the 'in-medium' meson mass variations in 
QM, QMVT and PQM, PQMVT models in order to see the influence of fermionic 
vacuum fluctuations. The sigma and pion  masses are calculated by determining
the curvature of grand potential at the global minimum.
\be 
\label{eq:pdergrand}
 m_{\pi,i}^{2}(T, \mu) = \frac{\partial^2 \Omega (T, \mu)}{\partial \pi_{i} 
 \partial \pi_{i}} \bigg|_{min}
\ee 
\be 
\label{eq:sigdergrand}
 m_{\sigma}^{2}(T, \mu) = \frac{\partial^2 \Omega (T, \mu)}{\partial \sigma 
 \partial \sigma} \bigg|_{min}
\ee 
The left panel of Fig.~\ref{fig:QMaRQMmass} shows the temperature variations of
meson masses for $\mu=0$, $\mu =\mu_{CEP}$ and $\mu > \mu_{CEP}$ in QM model while
the right panel shows the corresponding variations in QMVT model. In the chiral symmetry
broken mesonic phase, the sigma mass always decreases with temperature. Sigma mass 
increases again at high temperatures signaling chiral symmetry restoration and it 
becomes degenerate with the increasing pion mass which does not vary much below the 
transition temperature. The degenerate meson masses increase linearly with
$T$ after the chiral symmetry restoration transition\cite{Schaefer:2006ds} has taken place.
The temperature variations of $m_{\sigma}$ and $m_{\pi}$ masses at $\mu=0$ in Fig.~\ref{fig:QMaRQMmass:b}
are significantly modified due to the presence of fermionic vacuum fluctuations in QMVT model. If we compare
these variations with the corresponding QM model temperature variations of masses in Fig.~\ref{fig:QMaRQMmass:a}, 
we find that the mass degeneration in $m_{\sigma}$ and $m_{\pi}$ at $\mu=0$ in Fig.~\ref{fig:QMaRQMmass:b} becomes 
very smooth and it takes place at a higher temperature. Since the  chiral crossover transition on the 
temperature axis  at $\mu=0$ is quite sharp and fast in QM model, the mass degeneration trend in $m_{\sigma}$ 
and $m_{\pi}$ is also quite sharp and fast in Fig.~\ref{fig:QMaRQMmass:a}. Fermionic vacuum fluctuations
make the chiral crossover at $\mu=0$ very smooth in QMVT model and this gets reflected also in the setting up of
a very smooth mass degeneration trend at $\mu=0$ in Fig.~\ref{fig:QMaRQMmass:b}. Long-wavelength fluctuations of 
the order parameter, characterize the second-order phase transitions. Since the chiral phase transition turns
second order at the CEP, the sigma meson mass must vanish at the CEP because the effective potential completely 
flattens in the radial direction. Thus the sigma meson mass  drops below the pion mass near the CEP. 
We know that the CEP in QMVT model gets shifted to a significantly higher chemical potential 
under the influence of fermionic vacuum fluctuation, hence the sigma mass becomes almost zero at  $\mu_{CEP}=299.35$ MeV
in QMVT model as shown in Fig.~\ref{fig:QMaRQMmass:b}. The sigma mass goes to zero only at $\mu_{CEP}=151.7$ MeV 
in Fig.~\ref{fig:QMaRQMmass:b} in QM model. The discontinuities in mass evolutions respectively in Fig.~\ref{fig:QMaRQMmass:a}
and Fig.~\ref{fig:QMaRQMmass:b} , signal a first-order phase transition at small temperatures when $\mu>\mu_{CEP}=305$ MeV in 
both the models QM as well as QMVT.

\begin{figure*}[!tbp]
\subfigure[]{
\label{fig:PQMaRPQMmass:a} 
\begin{minipage}[]{0.45\linewidth}
\centering \includegraphics[height=8.6cm,width=5.6cm,angle=-90]{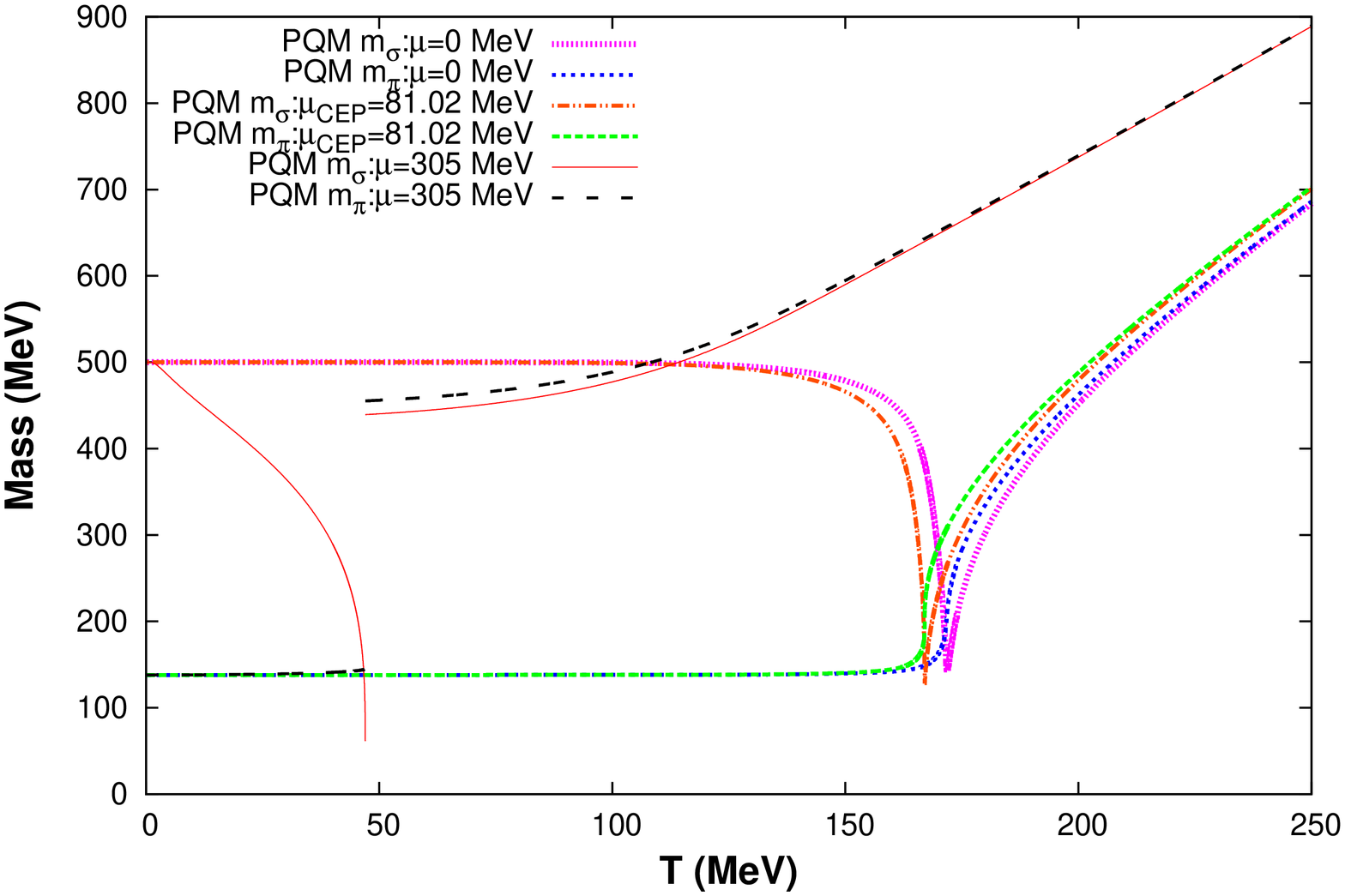}
\end{minipage}}%
\hspace{-0.03in}
\subfigure[]{
\label{fig:PQMaRPQMmass:b} 
\begin{minipage}[]{0.45\linewidth}
\centering \includegraphics[height=8.6cm,width=5.6cm,angle=-90]{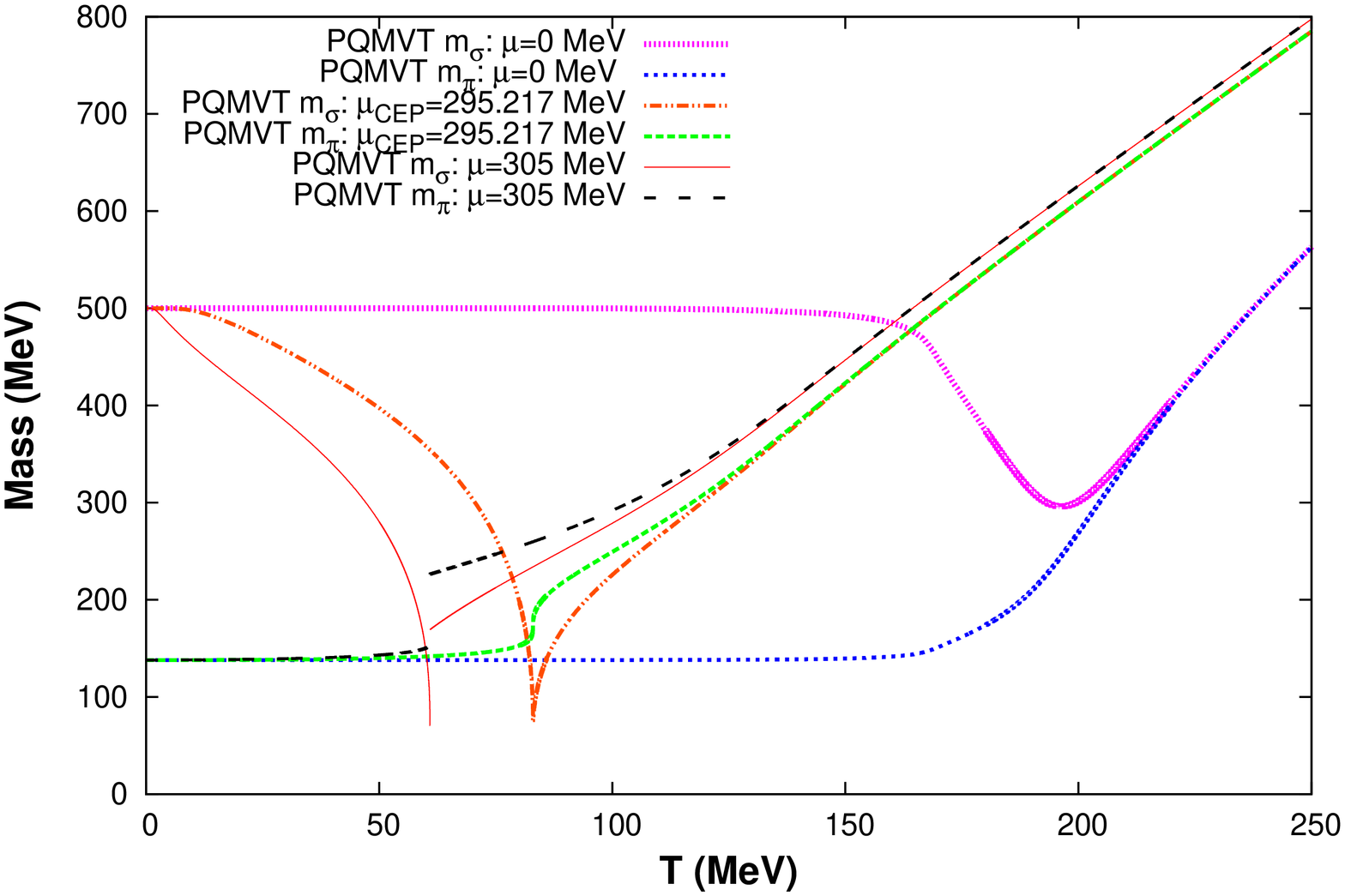}
\end{minipage}}
\caption{(a) Mass variations in the PQM model are shown in this Fig.
Two dotted lines in the rightmost part of the Fig. represent the
temperature variations of $m_{\sigma}$ and $m_{\pi}$  at $\mu=0$, the dash dotted
line in the middle part represents $m_{\sigma}$ while the line with small dash 
represents $m_{\pi}$ temperature variations at $\mu_{CEP}=81.02$ MeV and the solid
line in the leftmost part of the Fig. represents $m_{\sigma}$ while the line 
with dash represents $m_{\pi}$ temperature variations at $\mu=305$ MeV.
(b) Temperature variations of $m_{\sigma}$ and $m_{\pi}$ in the influence of 
fermionic vacuum fluctuations, have been shown for the PQMVT model calculations.
Lines represent the same  mass variations as in the left panel of the Fig. In PQMVT model
$\mu_{CEP}=295.217$ MeV.}
\label{fig:PQMaRPQMmass} 
\end{figure*}
Finally we will be investigating how the fermionic vacuum term influences the emergence of 
masss degeneration trend in $m_\sigma$ and $m_\pi$  as the chiral symmetry restoring transition 
takes place in the presence of Polyakov loop potential. Fig.~\ref{fig:PQMaRPQMmass:a} presents
temperature variations of meson masses at $\mu=0$, $\mu =\mu_{CEP}$ and $\mu > \mu_{CEP}$ in the
in PQM model calculations while Fig.~\ref{fig:PQMaRPQMmass:b} shows the corresponding mass variations 
in PQMVT model.
Since the  chiral crossover transition on the temperature axis  at $\mu=0$ becomes very sharp and rapid 
in the PQM model due to the influence of Polyakov loop potential, the mass degeneration trend in $m_{\sigma}$ 
and $m_{\pi}$ at $\mu=0$  also  becomes very sharp and fast in Fig.~\ref{fig:PQMaRPQMmass:a}. Here 
in the PQMVT model calculations also as in the case of QMVT model, the mass degeneration in $m_{\sigma}$ and $m_{\pi}$ at $\mu=0$ 
becomes very smooth in the influence of fermionic vacuum fluctuations as shown in Fig.~\ref{fig:PQMaRPQMmass:b}. Further this
mass degeneration takes place at a temperature which is highest of the corresponding mass degeneration temperatures in other models. 
This happens because, the chiral crossover transition at $\mu=0$ occurs at a temperature which is highest 
in the PQMVT model calculations, due to the combined effect of Polyakov loop potential and fermionic vacuum term. 
We point out that the sigma meson mass neither vanishes nor becomes very close to zero at CEP   
as shown  in Fig.~\ref{fig:PQMaRPQMmass:a} and Fig.~\ref{fig:PQMaRPQMmass:b} respectively for PQM and PQMVT model 
temperature variations. In PQM model, $m_{\sigma}$ in its temperature variation reaches the minimum value 
of 128 MeV as shown by the dash dotted line in Fig.~\ref{fig:PQMaRPQMmass:a} while in PQMVT model the minimum value reached
in the temperature variation of $m_{\sigma}$ is 76.0 MeV as evident from the dash dotted line in Fig.~\ref{fig:PQMaRPQMmass:b}.
It means that the effective potential  of PQM and PQMVT model does not completely flattens in the radial direction at the CEP.
The Polyakov loop expectation value is a scalar field which mixes up with the chiral order parameter and this effect 
hampers the flattening of the PQM model effective potential in the radial direction (i.e. the direction of $\sigma$ field)
at the CEP. In PQMVT model, this effect seems to be  considerably remedied by the presence of the fermionic vacuum term and  
the minimum value of $m_{\sigma}$ becomes 76.0 MeV in its temperature variation. Here also, the discontinuities in mass evolutions respectively in Fig.~\ref{fig:PQMaRPQMmass:a} and Fig.~\ref{fig:PQMaRPQMmass:b} ,signal a first-order phase transition at small temperatures when $\mu>\mu_{CEP}=305$ MeV in both the models PQM as well as PQMVT.

\section{Summary and Conclusion }
\label{sec:smry}
			In the beginning of the present work, we computed  the phase diagrams and pin-pointed the CEP
positions in the $\mu$  and T plane of all the four two quark flavor models QMVT,PQMVT,QM and PQM for the real life 
explicit chiral symmetry breaking with the experimental value of pion mass. We obtained the phase diagrams
with zero pion mass  also for the chiral limit in the QMVT and PQMVT models and located the respective positions of TCP.  
Since the presence or absence of TCP  in the phase diagram and its distance from CEP in a model calculation, 
influences the nature of critical fluctuations around CEP, we quantified the proximity of TCP to the CEP in 
the phase diagrams obtained for QMVT and PQMVT model calculations. 

The QMVT model CEP is positioned at $\mu_{CEP}$=299.35 MeV and $T_{CEP}$=32.24 MeV and it shifts to higher temperature 
$T_{CEP}$=83.0 MeV and lower chemical potential $\mu_{CEP}$=295.217 MeV in PQMVT model due to the effect of Polyakov loop. 
In QM model, the CEP is located at $T_{CEP}$=102.09 MeV and $\mu_{CEP}$=151.7 MeV and again in the influence of
Polyakov loop potential, it shows considerable shift  towards the temperature axis in PQM model and gets located 
at $T_{CEP}$=166.88 MeV and $\mu_{CEP}$=81.02 MeV. If we compare the location of CEP in QM and PQM models to the 
location of CEP in QMVT and PQMVT models, we find a considerably significant shift of CEP to large chemical 
potential and small temperature values for QMVT and PQMVT models due to the robust influence of fermionic vacuum term 
presence in the effective potential. In chiral limit of PQMVT model, the tricritical point(TCP) exists at 
$T_{t}$=137.09 MeV and $\mu_{t}$=240.14 MeV and the TCP in QMVT model is found at  $T_{t}$=69.06 MeV and 
$\mu_{t}$=263.0 MeV. The proximity of TCP to the CEP has been quantified by plotting the constant normalized
quark-number susceptibility ($R_{q}$=2) contours around CEP in the phase diagrams of PQMVT and QMVT models.
The second cumulant of the net quark number fluctuations on these contours is double to that of the free quark 
gas value and such enhancements are the signatures of CEP for the heavy-ion collision experiments. The TCP location 
is quite well inside the $R_{q}$=2 contour on the phase diagram of both the models QMVT as well as PQMVT.

 The CEP in phase diagrams is pin-pointed by tracking down the divergence in the quark number susceptibilities 
and scalar susceptibilities which show significant enhancement in a region around the CEP in the $\mu$ and T plane in 
comparison to their respective values for the free quark gas. In order to determine the shape of the critical 
region around CEP, we have plotted different contour regions  having different constant values of properly normalized 
quark number susceptibility ratio ($R_{q}$)  and scalar susceptibility ratio ($R_{S}$).
The different shapes of these contours as calculated in various models, throw light on the nature of criticality around CEP 
in those models. We have plotted the contours with three different values of quark number susceptibility ratio $R_q=2,3$ and 5,
in the $\mu$ and T plane relative to the CEP. If we compare the contours obtained in the PQM model to the contours
in pure QM model, we conclude that the presence of Polyakov loop potential, compresses the critical region
particularly in the T direction. Since the chiral crossover transition becomes faster and sharper due to the 
presence of Polyakov loop potential, the critical region in the T direction gets significantly compressed.
The analysis of the shape of $R_q=2,3$ and 5 contours in the QMVT and PQMVT models, tells us that
the  size of critical region is increased in a direction perpendicular to the crossover line
due to the influence of the fermionic vacuum fluctuations. This effect is less pronounced in PQMVT model
because of the compression of critical region width due to the presence of Polyakov loop potential 
while the QMVT model contours show a robust increase in the width of the critical region
in perpendicular direction to the crossover transition line . However, the extent and size of critical region 
in the PQMVT model  is noticeably larger in both the directions $\mu$ as well as T compared to that of  QMVT model
results. In the presence of fermionic vacuum term, CEP gets located at larger chemical potentials in
QMVT/PQMVT models. Since the quark determinant gets modified mostly at moderate chemical potentials by
the presence of Polyakov loop potential and further in its influence, the PQMVT model CEP shifts to a 
higher critical temperature when compared to the CEP in QMVT model, we obtain an enhancement of the 
critical region in PQMVT model. 

 We have plotted three contours around the CEP in PQM and QM models also for the normalized scalar susceptibility
ratios $R_{s}=10,15$ and 25. The shape of $R_{s}=10$ contour in PQM model is compressed in comparison to the 
pure QM model contours. The $R_{s}=25$ contour is not obtained in PQM model calculation because the minimum value
of $\sigma$ meson mass does not fall below 100 MeV. Since the value of $m_{\sigma}$ falls very rapidly and 
sharply from 500 MeV to 128 MeV, we obtain a very thin and small contour region even for $R_{s}=15$.
We get all the contours regions for $R_{s}=10,15$ and 25 with well defined size in the QM model calculations
because the $m_{\sigma}$ variation is smoother and slower in comparison to the corresponding PQM model
results and further the minimum in the $m_{\sigma}$ variation approaches almost zero value in QM model. 
Further we do not find the closure of $R_{s}=10$ contour on the temperature axis at $\mu=0$ MeV in both the models
QM as well as PQM. We obtain quite well defined and closed contour regions for $R_{s}=10,15$ and 25 in QMVT and PQMVT
model calculations which again become broader in the  direction perpendicular to the crossover line due to the 
presence of fermionic  vacuum fluctuations. For scalar susceptibility also, the critical region gets elongated
in the phase diagram and $\chi_\sigma$ is enhanced in the direction parallel to the first-order transition line.
Here also the presence of Polyakov loop potential in the PQMVT model, leads to the compression in the width of
critical region around CEP.The fermionic vacuum fluctuations, make the chiral crossover transition very 
smooth while the Polyakov loop potential makes it sharper and faster and these opposite effects
give a typical shape to the quark number susceptibility contours in the  PQMVT model. Similar effects 
can be seen in the scalar susceptibility contours also. In the influence of fermionic 
vacuum fluctuations only, the $\chi_\sigma$ contours in the pure QMVT model are broader and rounded. 

		In order to further investigate the nature of criticality in two flavour calculations,
we have numerically evaluated the critical exponents of the quark number 
susceptibility $\chi_q$ in QM,PQM,QMVT and PQMVT models. In these investigations,
the critical $\mu_{CEP}$ at fixed critical temperature $T_{CEP}$ is approached from 
the lower as well as higher $\mu$ sides in a path parallel to the $\mu$-axis in 
the ($T,\mu$)-plane. The calculation of the critical exponents has been done using
the linear logarithmic fit. If the $\mu_{CEP}$ is approached in QM model from the lower $\mu$ side,
we obtain a critical exponent equal to $\epsilon= 0.6379\pm 0.0002$  while the critical exponent 
is $\epsilon= 0.6648\pm 0.0001$ when the $\mu_{CEP}$ is approached from the higher $\mu$ side. 
The scaling starts around around $\log |\mu- \mu_{CEP} | < -.5$ in both the cases. These exponents 
show good agreement with the mean-field prediction $\epsilon = 2/3$. The influence of Polyakov
loop potential on the calculated values of critical exponents in PQM model, is negligible and we obtain
similar critical exponents as evaluated in QM model.

If the $\mu_{CEP}$ in QMVT model calculation is approached from the lower $\mu$ side, we obtain a 
critical exponent $\epsilon=m = 0.720\pm 0.00005$  which is larger in comparison of the corresponding 
critical exponent $\epsilon= 0.6379\pm 0.0002$  evaluated in the QM model. Due to the influence of fermionic
vacuum fluctuation, we find the presence of TCP in the chiral limit of QMVT model and it lies
quite well within the $R_{q}=2$ contour surrounding the CEP in the phase diagram. 
This larger critical exponent may be the consequence of the modification of criticality around CEP due to
the presence of TCP in its proximity. The scaling starts earlier when $\log |\mu- \mu_{CEP} | < 0.0$ and we observe
scaling over several orders of magnitude. We obtain smaller critical exponent $\epsilon=0.6938\pm 0.0001$
when the $\mu_{CEP}$ is approached from the higher $\mu$ side in PQMVT model.
It is pointed out that these exponents calculated in the presence of fermionic vacuum fluctuations
in the QM model are different from the mean-field prediction $\epsilon = 2/3$. The presence
of Polyakov loop compresses the width of critical region in PQMVT model but its effect on critical exponents 
is negligible. The critical exponents obtained in PQM model calculations are also similar to the critical exponents 
of QM model.

			Since the critical fluctuations are encoded in the variation of meson masses 
$m_\pi(T,\mu)$ and $m_\sigma (T,\mu)$ as one passes through the chiral symmetry restoring phase transition,
we have also investigated and compared the 'in-medium' meson mass temperature variations for 
$\mu=0$, $\mu =\mu_{CEP}$ and $\mu > \mu_{CEP}$  in QM, QMVT and PQM, PQMVT model calculations. 
If we compare the temperature variations of masses at $\mu=0$ in QM and QMVT model calculations,
we find that the mass degeneration in $m_{\sigma}$ and $m_{\pi}$ at $\mu=0$  becomes 
very smooth in QMVT model and it takes place at a higher temperature. Since the sharper and faster
chiral crossover transition occurring on the temperature axis  at $\mu=0$ in QM model, becomes very smooth in QMVT model
under the influence of fermionic vacuum fluctuations, the sharper mass degeneration trend in $m_{\sigma}$ 
and $m_{\pi}$ in QM model also becomes a very smooth mass degeneration trend in QMVT model. The sigma meson mass 
must vanish at the CEP since the chiral phase transition turns second order at this point and the effective 
potential completely flattens in the radial direction. In our QM and  QMVT model calculations, we have 
shown that the sigma meson mass becomes almost zero at $\mu=\mu_{CEP}$. in both the models. It has also been shown that
the discontinuities in mass evolutions, signal a first-order phase transition at small temperatures 
when $\mu>\mu_{CEP}=305$ MeV in both the models QM as well as QMVT. 

Finally our investigation gets concluded by the study of the influence of Polyakov loop potential 
on the emergence of masss degeneration trend in $m_\sigma$ and $m_\pi$  as the chiral symmetry 
restoration takes place in the presence of fermionic vacuum fluctuation.
Since the  chiral crossover transition on the temperature axis  at $\mu=0$ becomes very sharp and rapid 
in the PQM model due to the influence of Polyakov loop potential, the mass degeneration trend in $m_{\sigma}$ 
and $m_{\pi}$ at$\mu=0$  also  becomes very sharp and fast in PQM model. Here 
in the PQMVT model calculations also as in the case of QMVT model, the mass degeneration in $m_{\sigma}$ and $m_{\pi}$ at $\mu=0$ 
becomes very smooth in the influence of fermionic vacuum fluctuations. Further this mass degeneration happens at a 
temperature which is highest of the corresponding mass degeneration temperatures in other models. 
In the PQMVT model calculations, we obtain highest temperature for chiral crossover transition on the 
temperature axis at $\mu=0$. The combined effect of Polyakov loop potential and fermionic vacuum term is responsible for this. 
We point out that the sigma meson mass neither vanishes nor becomes very close to zero at $\mu=\mu_{CEP}$   
in PQM and PQMVT model temperature variations. In PQM model, $m_{\sigma}$ in its temperature variation 
reaches the minimum value of 128.0 MeV while the minimum value reached in the temperature variation of $m_{\sigma}$
in PQMVT model is 76.0 MeV. It means that the effective potential  of PQM and PQMVT model does not completely
flattens in the radial direction at the CEP. The Polyakov loop expectation value is a scalar field which mixes
up with the chiral order parameter and this effect hampers the flattening of the PQM model effective potential
 in the radial direction (i.e. the direction of $\sigma field)$ at the CEP. In the PQMVT model, this effect seems
to be  considerably remedied by the presence the fermionic vacuum term and we get minimum in $m_{\sigma}$ 
temperature variation at 76.0 MeV. We have also shown the discontinuities in mass evolutions which signal a first-order phase
transition at small temperatures when $\mu>\mu_{CEP}=305$ MeV in both the models PQM as well as PQMVT.

%
\begin{acknowledgments}
Valuable suggestions together with computational helps by Rajarshi Ray during the completion 
of this work are specially acknowledged. I am also thankful to Krysztof Redlich for 
fruitful discussions during the visit to  ICPAQGP-2010 at Goa in India.  General physics 
discussions with Ajit Mohan Srivastava are very helpful. computational support of the 
computing facility which has been developed by the Nuclear Particle Physics group of the
Physics Department, Allahabad University under the Center of Advanced Studies(CAS) 
funding of UGC, India, is also acknowledged. 
\end{acknowledgments}




\end{document}